# Cerebral blood flow monitoring using a deep learning implementation of the two-layer DCS analytical model with a 512×512 SPAD array

**Mingliang Pan,[a] Chenxu Li,[a] Yuanzhe Zhang,[a] Alan Mollins,[a] Quan Wang,[a] Ahmet T. Erdogan,[b] Yuanyuan Hua,[b] Zhenya Zang,[a] Neil Finlayson,[b] Robert K. Henderson,[b] David Day-Uei Li[a,*]**

[a]University of Strathclyde, Department of Biomedical Engineering, Glasgow, UK

[b]University of Edinburgh, School of Engineering, Integrated Nano and Micro Systems (IMNS), Edinburgh, UK

**Abstract**

**Significance:** Multi-layer (two- and three-layer) diffuse correlation spectroscopy (DCS) models improve cerebral blood flow index (CBFi) measurement sensitivity and mitigate interference from extracerebral tissues. However, their reliance on multiple predefined parameters (e.g., layer thickness, optical properties) and high computational load limits their feasibility for real-time bedside monitoring.

**Aim:** To develop a fast, accurate DCS data processing method based on the two-layer DCS analytical model, enabling real-time cerebral perfusion monitoring with enhanced brain sensitivity.

**Approach:** We employed deep learning (DL) to accelerate DCS data processing. Unlike previous DCS networks trained on single-layer models, our network learns from the two-layer DCS analytical model, capturing extracerebral *vs.* cerebral dynamics. Realistic noise was estimated from subject-specific baseline measurements using a 512×512 SPAD array at a large source-detector separation (35 mm). The model was evaluated on test datasets simulated with a four-layer slab head model via Monte Carlo (MC) methods and compared against conventional single-exponential fitting and the two-layer analytical fitting. Two *in-vivo* physiological response tests were also conducted to assess the real-world performance.

**Results:** The proposed method bypasses traditional curve-fitting, achieved real-time monitoring of CBF changes at 35 mm separation for the first time with a DL approach. Validation on MC simulations shows superior accuracy in relative CBFi estimation (4.1% error vs. 12.7% for single-exponential fitting) and significantly enhanced CBFi sensitivity (86.5% vs. 57.7%). Although the two-layer analytical fitting offers optimal performance, it depends on strict assumptions and preconditions, and its computational complexity makes it time-consuming and unsuitable for real-time monitoring applications.

Additionally, our method minimizes the influence of superficial blood flow and 750-fold faster than single-exponential fitting in a realistic scenario. *In-vivo* tests further validated the method's ability to support real-time cerebral perfusion monitoring and pulsatile waveform recovery.

**Conclusions:** This study demonstrates that integrating DL with the two-layer DCS analytical model enables accurate, real-time cerebral perfusion monitoring without sacrificing depth sensitivity. The proposed method enhances CBFi sensitivity and recovery accuracy, supporting future deployment in bedside neuro-monitoring applications.

***David Day-Uei Li**, E-mail: David.li@strath.ac.uk

## 1 Introduction

Cerebral blood flow (CBF) is a critical biomarker for brain health and function, supporting cognitive and neurological processes[1]. Existing techniques, such as transcranial Doppler ultrasound (TCD) is non-invasive but relies on experienced operators and is limited to larger arteries[2]. Functional magnetic resonance imaging (fMRI) and positron emission tomography (PET), on the other hand, are bulky and provide only 'snapshot' observations; they are not suitable for bedside applications[3,4]. In contrast, diffuse correlation spectroscopy (DCS) offers non-invasive, continuous, high-temporal-resolution CBF index (CBFi) measurements at the bedside[5]. DCS analyzes light intensity fluctuations caused by red blood cell movement using the autocorrelation function (ACF), given by $g_2 = \langle I(t)I(t+\tau)\rangle / \langle I(t)\rangle^2$, to assess blood flow dynamics[6,7].

Over the past two decades, DCS has evolved from continuous-wave (CW-DCS) to time-domain (TD-DCS) and frequency-domain (FD-DCS) variants[8,9]. Analytical models for complex tissue structures have advanced from semi-infinite homogeneous medium to two-layer and the three-layer models[10,11]. These developments have expanded DCS applications in brain health evaluation, neurovascular studies, cancer diagnosis, and therapy evaluation[3].

Traditionally, DCS maps measured $g_2$ curves to the semi-infinite model, allowing real-time CBFi measurements[12]. However, this model underestimates CBFi and is susceptible to extracerebral layer interference[13]. Attempts to improve CBFi sensitivity by fitting $g_2$ at short correlation times with the semi-infinite analytical model or increasing the source-detector separation ($\rho$) often reduce signal-to-noise ratio (SNR) and lead to inaccurate CBFi estimates[14]. There is a trade-off between the detection depth and SNR[15]. While advanced DCS variants such as pathlength-resolved DCS or TD-DCS can mitigate this issue, their complexity limits widespread adoption[16,17]. Multichannel DCS (MDCS) improves SNR by an order of magnitude, enabling deeper blood flow measurement[18,19]. For example, the ATLAS SPAD sensor developed by the University of Edinburgh achieves $\rho$~5 cm[20]. In addition, the two-layer and three-layer analytical models enhance CBFi sensitivity and enable separation of cerebral and extracerebral blood flow but require multiple predefined parameters for fitting (e.g., layer thickness and optical properties). If these parameters are mis-specified, significant errors may occur[21–23]. Additionally, these models are computationally intensive, limiting real-time application[24]. Integrating MDCS with multi-layer analytical models could enable more accurate CBFi estimation. However, advanced data processing techniques are required to overcome the existing limitations.

Deep learning (DL) has emerged as an efficient technique for DCS data processing in multiple studies[25–29]. It has demonstrated improved speed and robustness compared to traditional curve-fitting approaches[24,25]. For example, Poon *et al.* achieved 23-fold speed improvement[25], and Li *et al.* showed LSTM could improve BFi accuracy[27]. Nakabayashi *et al.* did explore an LSTM to separate shallow *vs* deep flow in a two-layer flow phantom, underlining the community's interest in accounting for layers[30]. However, to date these DL models have been trained mostly on the semi-infinite homogeneous (single-layer) analytical data, Monte Carlo simulations, or phantom experimental data, which are not suitable for CBFi recovery. To our knowledge, no prior publication has utilized training data from the two-layer DCS analytical model. The two-layer analytical model can accurately recover CBFi and relative CBFi (rCBFi) than the semi-infinite model while requiring fewer parameters than the three-layer model[22], making it an ideal candidate for training dataset generation.

To overcome the limitations of traditional fitting and leverage the capabilities of the new ATLAS SPAD sensor[20], this paper presents the following innovations:

We incorporated a SPAD-DCS system with a DL model trained on datasets generated by the two-layer DCS analytical model for real-time CBFi estimation. We validated our method using simulated test datasets to assess CBFi waveform recovery, rCBFi estimation, and CBFi sensitivity to brain and scalp blood flow perturbations, comparing the results with traditional single-exponential and two-layer analytical fittings. Finally, we evaluated the capability of the DL-based DCS system to monitor breath-holding and digestion-induced physiological responses in a healthy adult at a large $\rho$ of 35 mm.

## 2 Methods

### *2.1 Theory of the two-layer analytical model*

The DCS theory is based on the correlation diffusion equation (CDE), derived from correlation transfer equation (CTE) under the standard diffusion assumption[6]. This derivation is analogous to the photon diffusion equation (PDE) from the radiative transfer equation (RTE) using the $P_N$ approximation[6,31]. The analogy between the CTE and RTE was firstly established by Ackerson *et al.*[32]. The CDE is expressed as:

$$\left(\frac{D}{v}\nabla^2 - \mu_a - \frac{1}{3}\mu_s' k_0^2 \alpha \langle \Delta r^2(\tau) \rangle\right) G_1(\mathbf{r},\tau) = -S(\mathbf{r}), \tag{1}$$

where $G_1(\mathbf{r},\tau) = \langle E(r,t)E^*(r,t+\tau)\rangle$ is the unnormalized electric field temporal ACF, $D = v/(3\mu_s')$ is the photon diffusion coefficient, $v$ is the speed of light in the medium, $\tau$ is the time lag, $k_0 = 2\pi n_0/\lambda$ is the wavenumber of light in the scattering medium at the wavelength $\lambda$ and $n_0$ is the tissue refractive index, $\mu_a$ is the absorption coefficient, $\mu_s'$ is the reduced scattering coefficient, $S(r)$ is the source, and $\langle \Delta r^2(\tau)\rangle$ is the mean square displacement of scatterers. For diffusive motions, $\langle \Delta r^2(\tau)\rangle = 6D_b\tau$, where $D_b$ is the effective Brownian diffusion coefficient. In most practical applications, the Brownian motion model is accurate to describe the scatterers' motions[23,33,34]. The product $\alpha D_b$ is defined as BFi[4,7], where $\alpha$ is defined as the ratio of moving scatterers to total scatterers, assumed to be 1 in our simulations[35].

Following the analytical derivation process developed by Gagnon *et al.*[10], we assume an isotropic source incident at depth $z_0 = 1/(\mu_{a,1} + \mu_{s,1}')$, and scatters in each layer present independent Brownian diffusion motion. Then the CDE will be

$$\begin{aligned}
\left[D_1\nabla^2 - \mu_{a,1} - 2\mu_{s,1}' k_0^2 D_{b,1}\tau\right] G_{1,1}(x,y,z,\tau) &= -\delta(x,y,z-z_0) \quad 0 \le z \le l, \\
\left[D_2\nabla^2 - \mu_{a,2} - 2\mu_{s,2}' k_0^2 D_{b,2}\tau\right] G_{1,2}(x,y,z,\tau) &= 0 \quad\quad l \le z,
\end{aligned} \tag{2}$$

where $j = 1,2$ refers to the layer indices, $G_{1,j}$, $D_j$, $\mu_{a,j}$, $\mu_{s,j}'$, and $D_{b,j}$ are the unnormalized electric field temporal ACF, diffusion coefficient, absorption coefficient, reduced scattering coefficient, and Brownian diffusion coefficient in Layer $j$, respectively, $l$ is the thickness of Layer 1. The Fourier domain solution to Eq. (2) at the surface of Layer 1 is

$$\tilde{G}_{1,1}(s,z,\tau) = \frac{\sinh[\alpha_1(z_b+z_0)]}{D_1\alpha_1} \times \frac{D_1\alpha_1\cosh[\alpha_1(l-z)] + D_2\alpha_2\sinh[\alpha_1(l-z)]}{D_1\alpha_1\cosh[\alpha_1(l+z_b)] + D_2\alpha_2\sinh[\alpha_1(l+z_b)]} - \frac{\sinh[\alpha_1(z_0-z)]}{D_1\alpha_1}, \quad (3)$$

where $\alpha_j^2 = \left(D_j s^2 + \mu_{a,j} + 2v\mu'_{s,j}k_0^2 D_{b,j}\right)/D_j$, $v$ is the light speed, $z_b = 2D_1(1+\mathrm{R_{eff}})/(1-\mathrm{R_{eff}})$. The Fourier inversion of Eq. (3) is

$$G_1^1(\rho,z,\tau) = \frac{1}{2\pi}\int_0^\infty \tilde{G}_{1,1}(s,z,\tau)sJ_0(s\rho)\mathrm{d}s\,, \quad (4)$$

where $J_0$ is the zeroth order Bessel function of the first kind[10,36].

The normalized electric field temporal ACF, $g_1(\tau)$, is related to the normalized light intensity ACF, $g_2(\tau)$, through the Siegert equation[37]:

$$g_2(\tau) = 1 + \beta|g_1(\tau)|^2, \quad (5)$$

where $\beta$ depends on the laser stability, coherence length, and the number of speckles detected[34]. The experimentally measured light intensity ACF can be calculated as:

$$g_2(\tau) = \frac{\langle I(t)I(t+\tau)\rangle}{\langle I(t)\rangle^2}, \quad (6)$$

where $\langle\ldots\rangle$ denotes the average over the integration time $T_{int}$, and $I(t)$ is the measured light intensity fluctuation. By fitting the measured light intensity ACF to the analytical solution, the BFi ($\alpha D_b$) can be extracted.

*2.2 Training dataset preparation*

The clean training dataset was generated using the two-layer DCS analytical model. Based on previous studies[10,38], we varied only the dominant parameters of the model. Specifically, we varied the brain $\mu_a$ and $\mu_s'$, the extracerebral layer thickness, and the extracerebral layer $D_b$ ($D_{b_extra}$) and brain $D_b$ ($D_{b_brain}$), while keeping other parameters constant. The brain $\mu_a$ varied linearly from 0.005 to 0.025 mm$^{-1}$ with a step size of 0.005 mm$^{-1}$, and the brain $\mu_s'$ varied from 0.9 to 1.3 mm$^{-1}$ (the step size: 0.1 mm$^{-1}$). The extracerebral layer thickness varied from 8 to 15 mm (the step size: 1 mm). $D_{b_brain}$ varied linearly from $5\times10^{-7}$ to $5\times10^{-5}$ mm$^2$/s with a step of $5\times10^{-7}$ mm$^2$/s (100 steps). Each $D_{b_brain}$ corresponded to 20 steps of $D_{b_extra}$, with the extracerebral-to-brain $D_b$ fraction ranging from 0.05 to 0.3. This relationship was estimated from previous reports[39,40]. In total, we simulated 400,000 clean $g_2$ data (5×5×8×100×20). The $D_b$, physiological and optical properties for the two-layer analytical model were adopted from previous studies[10,14,38,39,41–43], to ensure a thorough coverage of relevant ranges. $\rho$ was fixed at 35 mm, $\beta$ was assumed to be 0.5, and the time lag $\tau$ ranged from 1.28 to 39.68 μs (step size: 1.28 μs) to align with the hardware settings (see Sec. 2.3). We fixed $\beta$ as a constant because scaling the $g_2$ curve to the same range will eliminate the influence of $\beta$ variance at a given flow rate. As a result, it is unnecessary to vary $\beta$ to account for potential hardware instability (see Sec. 2.3 for details). The parameter configurations are summarized in Table 1.

**Table 1** Simulation parameters for the two-layer analytical model.

| Tissue type | $\mu_a$ (mm$^{-1}$) | $\mu_s'$ (mm$^{-1}$) | Tissue thickness (mm) | $D_b$ (mm$^2$/s) |
|---|---|---|---|---|
| Extracerebral layer | 0.019 | 0.86 | (8, 15) | (0.05, 0.3) × $D_{b_brain}$ |
| Brain | (0.005, 0.025) | (0.9, 1.3) | ∞ | ($5\times10^{-7}$, $5\times10^{-5}$) |

### *2.3 SPAD-DCS system and noise calculation*

As reported earlier[20], the SPAD sensor ATLAS, with embedded on-chip autocorrelation computation optimized for DCS applications, has demonstrated deep and high-speed CBF monitoring. In this work, we operated ATLAS in the ensemble DCS mode, where all 128×128 macropixels (each composed of 4×4 micropixels) were combined and averaged to output 31 time lags of the light intensity ACF. The pixel clock (*PixClk*) was set to 25 MHz, corresponding to a time lag range of 1.28–39.68 µs, enabling deeper/faster flow information capture. The iteration number was set to 4096, corresponding to an integration time of 5.24 ms. We used a continuous-wave laser source (DL785-100-S, 785 nm, 100 mW, CrystaLaser) coupled with a multimode optical fiber (MMF, M143L01, Ø600 $\mu m$, 0.22 NA, Thorlabs) to illuminate tissues. The detector fiber (MMF, M59L01, Ø1000 $\mu m$, 0.50 NA, Thorlabs) tip was placed 23 mm from the SPAD chip, the optimal distance for maximizing speckle contrast[20]. Both fibers were held by a custom 3D-printed probe, maintaining a 35 mm separation. The overall system setup is illustrated in Fig. 1. We processed SPAD ACF data using single-exponential fitting, a simplified model commonly preferred for real-time measurements. The analytical normalized light intensity ACF can be simplified to a single-exponential decay function at a small $\tau$ range[44,45]:

$$g_2(\tau) = 1 + \beta e^{-2\tau/\tau_c}, \quad (7)$$

where $\tau_c$ is the decorrelation time. The reciprocal of $\tau_c$, known as the decorrelation speed, is directly proportional to blood flow rate and can be used to quantify blood flow changes[13]. We simplified Eq. (7) to the following format to facilitate data analysis:

$$g_2(\tau) = a + be^{-c\tau}, \quad (8)$$

The validity of this simplification is demonstrated in Fig. S3 of the Supplementary Material. In the following sections, we compare cerebral perfusion measured by DL and single-exponential fitting method through comparing rCBFi measured with the DL model and the relative change in single-exponential fitting-recovered decorrelation speed.

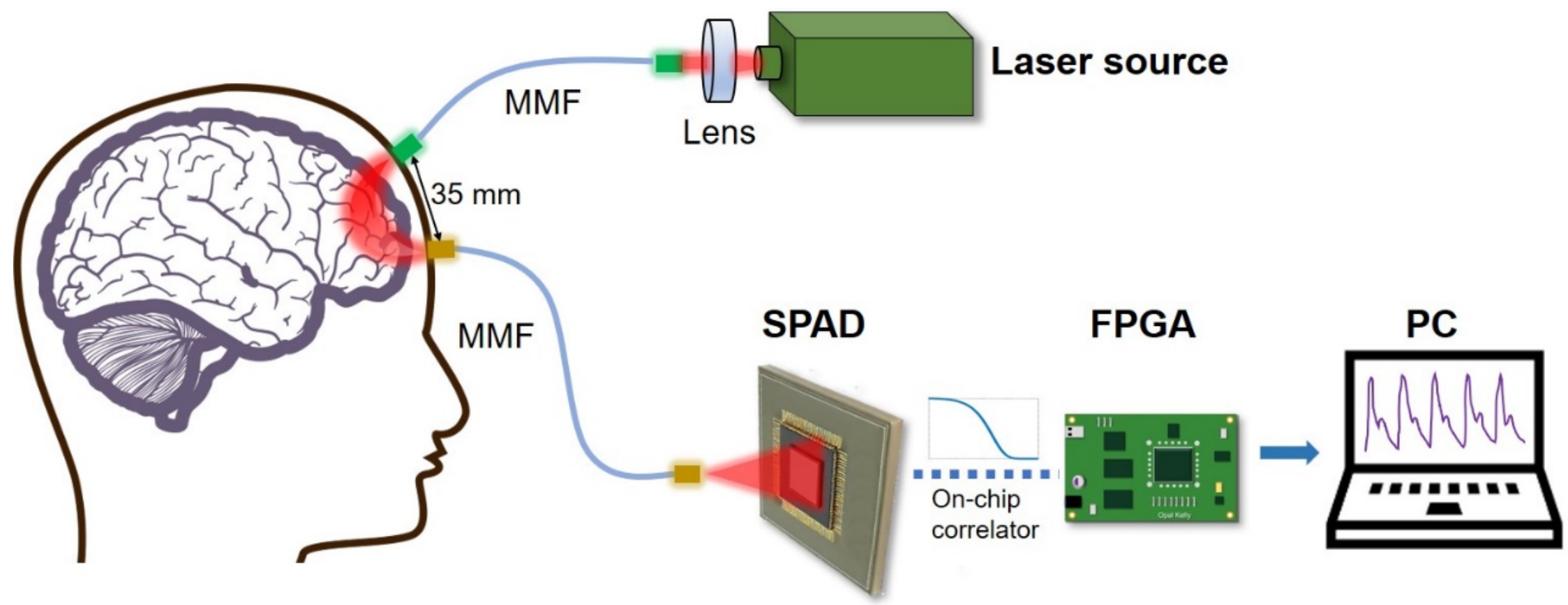

**Fig. 1** The SPAD-DCS system setup. The 785 nm continuous wave laser was coupled into a multi-mode fiber (MMF, M143L01, Ø600 $\mu m$, 0.22 NA, Thorlabs), the scattered photons was collected by another MMF (M59L01, Ø1000 $\mu m$, 0.50 NA, Thorlabs). The output of the SPAD array was received by an Opal Kelly FPGA board (XEM7310-A200) and transferred to the PC through a USB 3.0 cable. The on-chip computed ACF is sent to the trained DL model for real-time CBFi/rCBFi display.

We applied this system to a healthy human adult (male, 29 years old) at the resting-state to collect the baseline autocorrelation. The probe was secured to the participant's forehead using an adjustable Velcro strap to ensure stability and comfort. The test was repeated five times, recording 1000 ACF frames per session, yielding a total of 5,000 baseline samples for noise estimation. We use $X(\tau)$ to represent the collected ACF data. To prevent overestimation of noise due to pulsatile CBF fluctuations, we applied a multi-step correction process as illustrated in Fig. 2 (block with light blue background). First, $X(\tau)$ was scaled to [1, 1.5] using equation:

$$x(\tau) = \frac{X(\tau) - \min\big(X(\tau)\big)}{\max\big(X(\tau)\big) - \min\big(X(\tau)\big)} \times 0.5 + 1, \tag{9}$$

where $x(\tau)$ is the scaled measured ACF. Second, $x(\tau)$ was fitted with a single-exponential decay function, $f(\tau) = a + be^{c\tau}$. Third, we calculated the standard deviation of the residual (as a function of $\tau$), i.e., $\sigma = std\big(x(\tau) - f(\tau)\big)$, *std()* is the standard deviation calculation function used in MATLAB. We obtained five sequences of $\sigma$ from the five tests and calculated their mean and standard deviation. The averaged sequence was then scaled by ±30% to encompass the variability observed across the tests (actual range is from -27.8% to +30.1% relative to the averaged sequence). This yielded three levels of $\sigma$: the averaged $\sigma$, +30% and –30% from mean. Finally, the three levels of $\sigma$ are substituted into a Gaussian distribution model with zero mean to generate noise and added to the simulated clean dataset[44]. In total, we generated 1,200,000 training data samples (through combination of 400k clean curves × 3 noise levels). Notably, we rescaled each noise-added curve with the same method (Eq. 9) prior to training. Likewise, any experimental SPAD ACF is scaled with Eq. (9) before input to the model, to maintain consistency between training and inference. The work flow of our method is illustrated in Fig. 2, including five parts: noise calculation, training dataset generation, test dataset generation, model training, and regression. The detailed ACF data processing for training and testing our model is shown in Fig. 3.

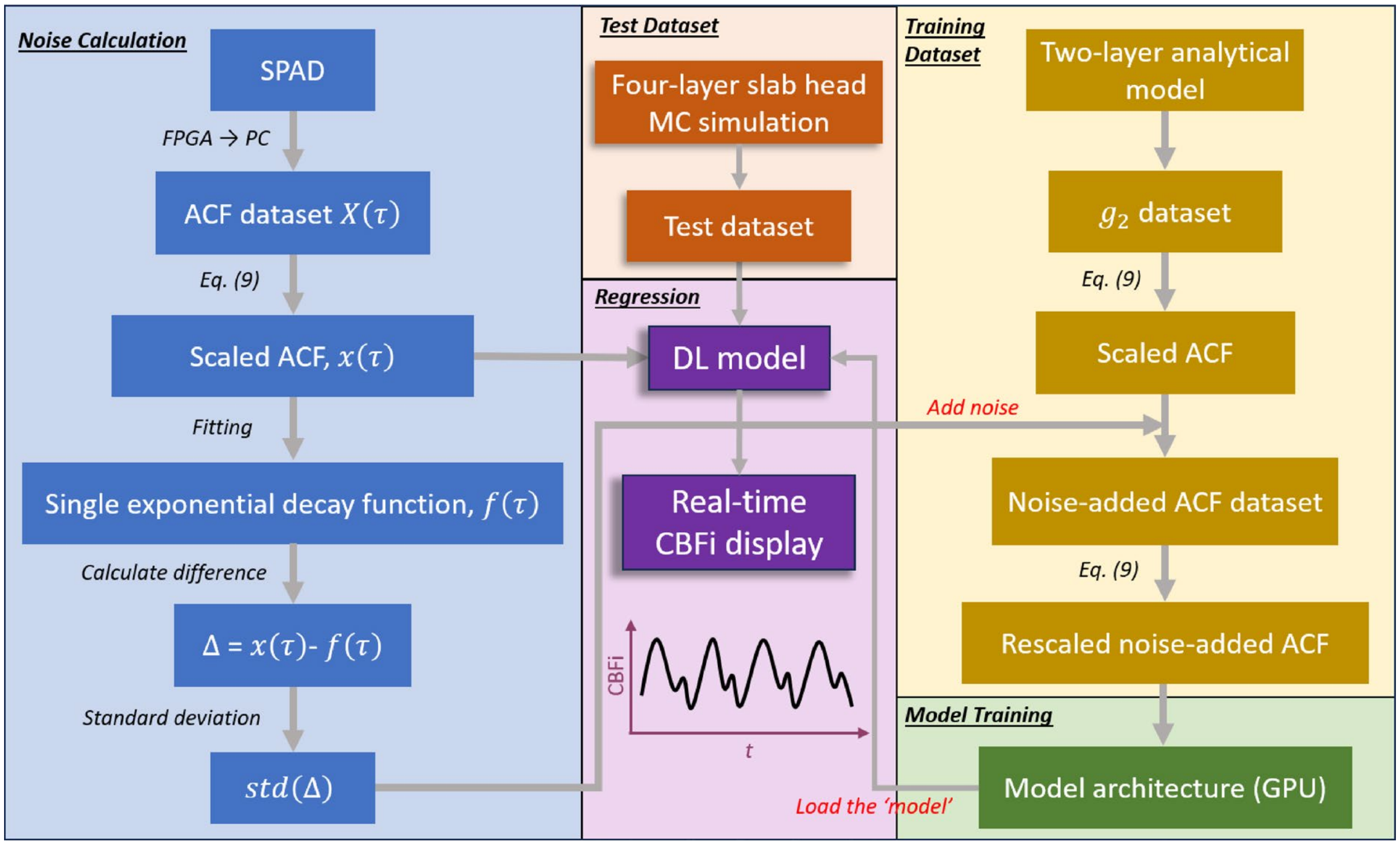

**Fig. 2** Flowchart of the data processing protocol. The process consists of five main components: (1) the 'Noise Calculation' block, which calculates the standard deviation from experimental baseline data; (2) the 'Dataset Generation' block, which outlines the procedure for generating clean data and adding noise; (3) the 'Model Training' block, which describes the training of the model on a GPU; (4) the 'Test Dataset' block, which illustrates the test dataset generation using MC simulation to assess our model, and (5) the 'Regression' block, which represents real-time CBFi/rCBFi display using the trained model and the preparation of experimental data before input to the model.

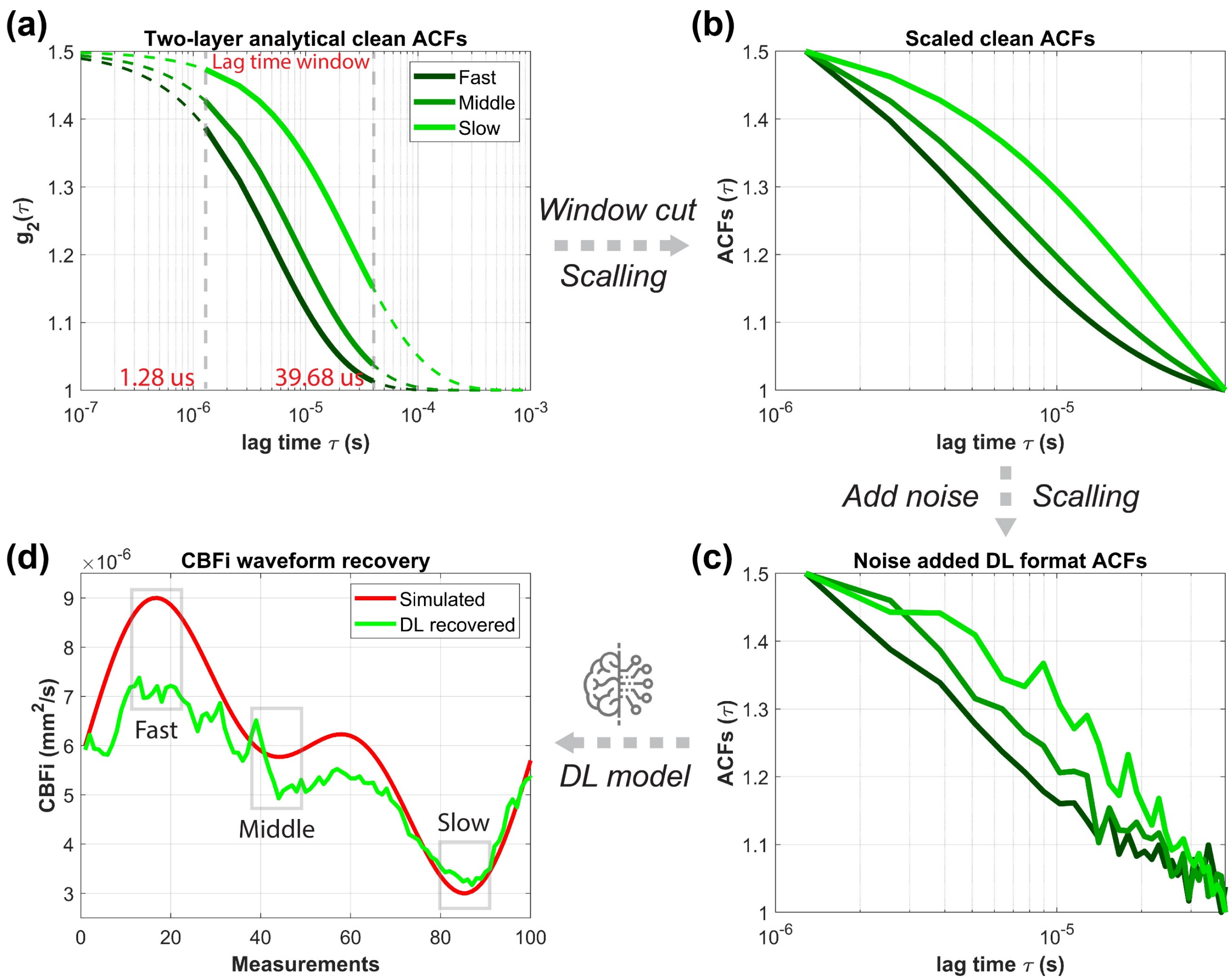

**Fig. 3** The ACF data processing workflow for training and testing our DL model. (a) ACFs were generated using the two-layer analytical model at different flow rates (fast, medium, and slow as the representatives). Each curve was truncated according to the time lag window used in the SPAD-DCS system. (b) The windowed ACFs were scaled to [1, 1.5]. (c) Experimentally derived noise was added to the scaled ACFs, and the noisy ACFs were rescaled to [1, 1.5]. These processed ACFs were then formatted for input into the DL model. (d) Representative CBFi waveforms recovered by the DL model from the test dataset.

### *2.4 Deep learning model architecture*

In this study, we employed a Long Short-Term Memory (LSTM) network as our deep learning model architecture. LSTM, an advanced variant of the RNN architecture designed for model sequential data, has been applied in several studies for DCS data analysis[30,46]. Our LSTM model architecture is illustrated in Fig. 4. The dataset consists of 1,200,000 samples, with 80% used for training (960,000 samples) and 20% for validation. CBFi was used as the training label, and each value was scaled by $10^6$ to prevent slow training convergence. We chose an LSTM with 2×128 units as it offered a good balance of complexity and performance, and similar RNN-based models have proven effective in DCS analysis[27,28,30]. Based on our model architecture (Table 2), the total number of trainable parameters is 198,273, yielding a training sample-to-parameter ratio of 4.84:1. This ratio supports effective generalization while reducing the risk of overfitting, which is generally sufficient to avoid overfitting. The model was trained by minimizing the mean squared error (MSE) loss function, with Adam as the optimizer. To avoid overfitting, we applied dropout rate of 0.3 and L2 regularization (weight decay of $10^{-4}$, batch size: 256 and epochs: 1000) during the training. A fully connected layer processes the unit's output to regress CBFi. The model was developed using Pytorch and executed on our workstation (GPU: NVIDIA Quadro RTX 5000 with 16 GB memory). The training and validation loss curves are shown in Fig. S2 (Supplementary Material).

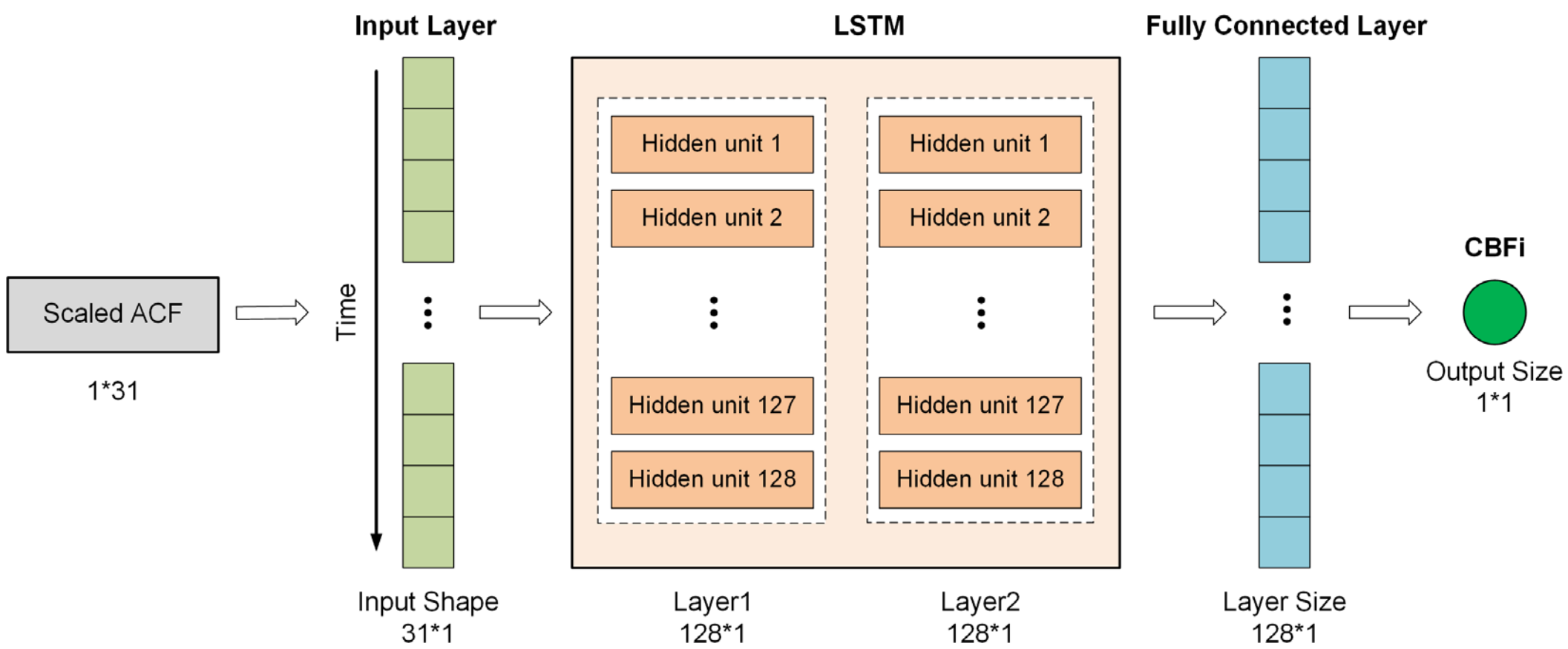

**Fig. 4** The proposed LSTM model architecture.

**Table 2** LSTM architecture parameters.

| Parameters | Values |
|---|---|
| Input size | 31×1 |
| Number of hidden layers | 2 |
| Layer unit | 128 |
| Loss function | MSE |
| Optimizer | Adam |
| Learning rate | 0.0001 |
| Batch size | 256 |
| Epoch | 1,000 |
| Output size | 1 |

### *2.5 Monte Carlo simulation for test dataset generation*

We conducted MC simulations using the voxel-based Monte Carlo eXtreme (MCX)[47] toolkit in MATLAB (R2023b, The MathWorks) to generate the test dataset. We simulated a slab human head model with a volume of $200\times200\times100$ mm$^3$, segmented into four layers, each layer representing scalp, skull, cerebrospinal fluid (CSF), and brain tissues. 785 nm light was used in all MC simulations, the layer thicknesses and optical parameters at 785 nm are listed in Table 3. Each simulation was executed with $5\times10^8$ photons and the detector was positioned at $\rho$ = 35 mm to record the photon transfer and photon pathlength, thereby enabling the calculation of the temporal light field ACF[7,48]:

$$G_1(\tau) = \frac{1}{N_p}\sum_{n=1}^{N_p} \exp\left(\sum_{i=1}^{N_{tissue}} -\frac{1}{3} Y_{n,i} k_0^2 \langle \Delta r^2(\tau) \rangle_i \right) \exp\left(-\sum_{i=1}^{N_{tissue}} \mu_{a,i} L_{n,i}\right), \quad (10)$$

where $N_p$ is the number of detected photons at each detector, $N_{tissue}$ is the number of tissue types ($N_{tissue}$ = 4 in our case), $Y_{n,i}$ and $L_{n,i}$ are the total momentum transfer and total pathlength of photon $n$ in Layer $i$, and $\mu_{a,i}$ is the absorption coefficient in Layer $i$. $\langle \Delta r^2(\tau) \rangle_i = 6D_{bi}\tau$ is the mean square displacement of the scattered particles in Layer $i$, where $D_{Bi}$ is the effective Brownian diffusion coefficient in Layer $i$. The correlation delay time $\tau$ was set between 1.28 µs and 39.68 µs with 31 linearly spaced data points to agree with the SPAD output data format. $\beta$ was set to 0.5, $g_2(\tau)$ curves were obtained by substituting the normalized electric field ACF into Eq. (5). The anisotropic factor $g$ was set to 0.89 for all MC simulations.

**Table 3** Baseline flow, physiological and optical parameters at 785 nm of the four-layer model simulation.

| | Layer | Layer thickness (mm) | $\mu_a$ (mm$^{-1}$) | $\mu_s'$ (mm$^{-1}$) | $D_b$ (mm$^2$/s) |
|---|---|---|---|---|---|
| **four-layer slab** | Scalp | 3 | 0.019 | 0.726 | $1 \times 10^{-6}$ |
| | Skull | 7 | 0.014 | 0.946 | $8 \times 10^{-8}$ |
| | CSF | 2 | 0.001 | 0.002 | $1 \times 10^{-8}$ |
| | Brain | ∞ | 0.020 | 1.210 | $6 \times 10^{-6}$ |

We use $D_{b_brain}$ and $D_{b_scalp}$ to represent the Brownian diffusion coefficients in brain and scalp, respectively. We first simulated a pulsatile waveform of $D_{b_brain}$ while gradually increasing $D_{b_scalp}$ to show the model's performance in recovering the $D_{b_brain}$ waveform as well as its ability to identify the influence of $D_{b_scalp}$. For the simulations, $D_{b_brain}$ ranges from $3\times10^{-6}$ mm$^2$/s to $9\times10^{-6}$ mm$^2$/s follows a pulsatile pattern, while the $D_{b_scalp}$ gradually increases from $5\times10^{-7}$ mm$^2$/s to $1.5\times10^{-6}$ mm$^2$/s. Other parameters remain the same as in the baseline condition, and one level of noise (middle) was added to the simulated data as described previously. For $\text{CBFi}_{\text{baseline}}$ recovery, we simulated 1000 noise-added data samples under the baseline condition (Table 3). The mean values of the recovered CBFi from the DL model and the two-layer analytical fitting, the decorrelation speed measured using single-exponential fitting across the 1000 data samples were used as the baseline for calculating relative cerebral perfusion changes, respectively. rCBFi was calculated as rCBFi $= \text{CBFi}/\text{CBFi}_{\text{baseline}}$[7,27].

To quantify CBFi sensitivity, $D_{b_brain}$ was varied by ±25% and ±50% relative to the baseline, while maintaining $D_b$ in other layers constant. Similarly, $D_{b_scalp}$ was varied by ±25% and ±50% relative to the baseline while keeping $D_b$ in other layers constant, allowing us to evaluate the model's CBFi measurement sensitivity to scalp BFi (SBFi) changes. At each perturbation level, 1000 noise-added datasets were generated. The sensitivity is defined as[49]:

$$S = \frac{\frac{(\text{CBFi} - \text{CBFi}_0)}{\text{CBFi}_0}}{\frac{(D_b - D_{b0})}{D_{b0}}} \times 100\%, \quad (11)$$

where $\mathrm{CBFi}$ and $\mathrm{CBFi}_0$ represent the recovered CBFi under perturbed and baseline conditions, respectively, $D_b$ and $D_{b0}$ denote the simulated brain or scalp Brownian coefficient at perturbation and baseline conditions. For single-exponential fitting, the numerator of Eq. (11) is replaced by the change in decorrelation speed. Ideally, a measured CBFi sensitivity close to 100% is preferred, as it indicates accurate detection of cerebral blood flow changes. Conversely, a measured CBFi sensitivity to SBFi changes close to 0% suggests that the model is robust against SBFi variations, minimizing confounding effects from extracerebral blood flow.

We also calculated the recovered rCBF error by the DL model, two-layer analytical fitting and single-exponential fitting, calculated using below equation:

$$\epsilon = (\mathrm{rCBF} - rD_b)/rD_b \times 100\%, \tag{12}$$

where $\epsilon$ represents the percentage error in recovered relative flow change.

For comparison, we added noise to the test dataset used for two-layer analytical model fitting. The noise was generated using a widely accepted model from a previous study[44], employing a linear-$\tau$ sampling strategy[45]. Details of the noise model and the specific parameters used are provided in the Supplementary Material. Specifically, we assumed a photon count rate of 3.3 kHz at a source-detector separation of $\rho$ = 35 mm and an integration time of 180 s. This photon count rate was estimated based on values reported in a prior study[39] at $\rho$=30 mm and derived using the relationship of the unnormalized field autocorrelation function at zero time lag, $G_1(0)$[57,58]. The noise level is shown in Fig. 5(d). For each perturbation level, we generated 1,000 noisy ACFs for the two-layer analytical fitting. The dynamic parameters $D_b$ in the brain and extracerebral layers were treated as fitting variables, while the coherence factor $\beta$ was assumed to be known and consistent with the value used in the four-layer model simulations (see Supplementary Material).

*2.6 Model's stability test*

To assess the model's stability, we generated a series of datasets by varying the brain layer's $\mu_a$, $\mu_s'$, and the scalp and skull thicknesses using MC simulations based on the four-layer head model. Specifically, each parameter was assigned five evenly spaced values: [0.005, 0.010, 0.015, 0.020, 0.025] mm$^{-1}$ for brain $\mu_a$; [0.9, 1.0, 1.1, 1.2, 1.3] mm$^{-1}$ for brain $\mu_s'$; [2, 3, 4, 5, 6] mm for scalp thickness ($L_{scalp}$); and [5, 6, 7, 8, 9] mm for skull thickness ($L_{skull}$). To compare the model's stability in CBFi recovery, the dynamic parameters $D_b$ across all groups are same as listed in Table 3. For rCBFi recovery, we simulated ±50% variations in $D_{b_brain}$ with a baseline value of $D_{b_brain}$= 6×10$^{-6}$ mm$^2$/s. The group with $\mu_a$= 0.015 mm$^{-1}$, $\mu_s'$= 1.1 mm$^{-1}$, $L_{scalp}$= 4 mm, and $L_{skull}$= 7 mm was set as the reference group for calculating CBFi and rCBFi errors, defined as

$$E = \left(p - p_{ref}\right)/p_{ref} \times 100\%, \tag{13}$$

where $p$ denotes the recovered CBFi or rCBFi for a given parameters group, and $p_{ref}$ represents the recovered CBFi or rCBFi in the reference group. Not that when evaluating a parameter's sensitivity, all other parameters were fixed at their reference values. The rCBFi recovery errors were calculated under two variations (-50% with $D_{b_brain}$= 3×10$^{-6}$ mm$^2$/s and +50% with $D_{b_brain}$= 9×10$^{-6}$ mm$^2$/s). The final reported rCBFi error for each group is defined as the mean of the two results, helping to generalize the findings and avoid bias associated with the direction of variation. We used the errors in CBFi and rCBFi to characterize the model's stability. An ideal model would yield errors close to zero, indicating consistent and reliable performance across parameter variations.

*2.7 Human physiological response tests*

We first applied the proposed DL model, integrated with the SPAD sensor, to evaluate cerebral perfusion changes during a simple physiological response paradigm, expiratory breath-holding, which is expected to induce an increase in CBFi. The experiment was conducted on a healthy 29-year-old male subject. Each test session consisted of three phases: 30 seconds of normal breathing, followed by 30 seconds of expiratory breath-holding, and concluding with 30 seconds of resumed normal breathing. To assess the stability of our method, the test was repeated on three separate days.

Moreover, we conducted an additional test involving a routine activity, eating lunch, to evaluate digestion-induced changes in cerebral perfusion. CBFi measurements were recorded 30 minutes before the meal and at 5, 30, 75, and 120 minutes post-meal. In each test phase, 5000 frames of autocorrelation data were collected. As illustrated in Fig. 1, a custom 3D-printed probe was used to hold the source and detector fibers, which were attached to the participant's forehead with an adjustable Velcro strap. During the tests, the power density from the source fiber tip was attenuated to remain below the Maximum Permissible Exposure limit set by the American Laser Safety Standard (28.5 mW for 785 nm laser)[50]. The participant wore laser safety goggles to prevent laser exposure to the eyes. Ethical approval was granted by the Biomedical Engineering Departmental Ethics Committee at the University of Strathclyde.

## 3 Results

*3.1 Noise characterization*

Fig. 5(a) shows an example of detected blood flow changes during a pulse cycle, recovered using single-exponential fitting. The fitting method is employed to illustrate how CBFi changes over a pulse cycle affect the measured signal. Fig. 5(b) presents the scaled measured ACF and the corresponding fitted decay curves at the peak and trough in Fig. 5(a). Fig. 5(c) displays the residuals, $\Delta = x(\tau) - f(\tau)$, calculated at the peak and trough at each of the 31 $\tau$. Fig. 5(d) is the standard deviation $\sigma$ of the residuals averaged and scaled across the five test phases (each phase has 5,000 frames of ACFs), which serves as the noise model for generating synthetic noise (see Sec. 2.2.).

We scaled the SPAD ACF data before noise characterization (using Eq. 9) to match the training input format. Because the SPAD's autocorrelator uses linear lag spacing, the noise standard deviation is expected to follow an exponential decay[45]. Indeed, in Fig. 5(d) the standard deviation is lower at the first and last lag points compared to the middle; we believe this occurs due to our scaling and the rapid decay of the $g_2$ curve over the initial lag range at high flow rates. Additionally, since the correlation time ranges from 1.28 μs to 39.68 μs, corresponding to the very beginning of a full $g_2$ decay, so the correlation values drop off very rapidly (especially for high flow). Consequently, the first point of the scaled SPAD ACF is 1.5 (by design of the scaling).

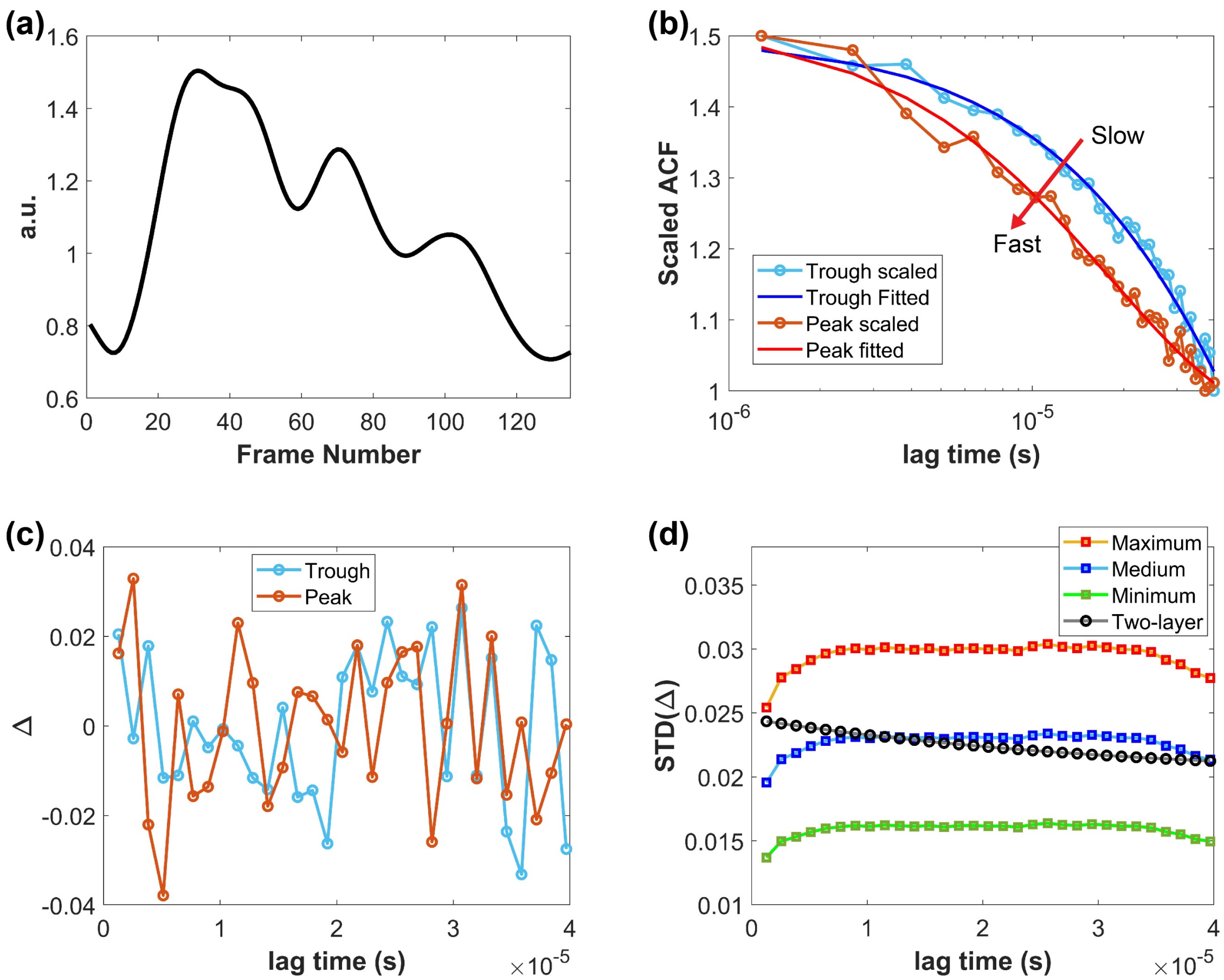

**Fig. 5** Examples of noise calculation. (a) A waveform of detected blood flow changes, where red and light blue circles indicate the peak and trough of one cycle. (b) Scaled SPAD ACF and the corresponding fitted single-exponential decay curves at the peak and trough. (c) The difference between the scaled SPAD ACF and the fitted single-exponential decay function at the peak and trough. (d) The standard deviation was calculated over 1000 frames across five tests, we scaled the original $\sigma$ to get three levels of noise added to the clean dataset. The gray line with black circles represents the standard deviation used to generate the noisy dataset for the two-layer analytical model fitting, details can be found in Supplementary Material.

### *3.2 CBFi waveform recovery on simulated data*

In this section, we firstly visualized the performance of the proposed DL model in CBFi waveform recovery and assessed its ability to isolate SBFi variations, in comparison with the two-layer analytical fitting. Fig. 6(a) and (b) show the presentative ACFs from the four-layer model simulation and used as input for the trained DL model. Fig. 6(c) presents the recovered CBFi waveform from the simulated dataset of the four-layer slab head model. The results show that the DL can effectively reconstruct CBFi waveforms and achieve accuracy comparable to the two-layer analytical fitting, although both methods tend to underestimate absolute CBFi values. This underestimation arises from the two-layer analytical model's bias, where the scalp and skull are grouped into a single extracerebral layer. Since blood flow in the skull is typically minimal, this grouping leads to a downward bias in CBFi estimation as well as SBFi estimations[22,38]. The results presented in Fig. 6 align with the two-layer analytical model-induced bias, suggesting that our model has learned similar characteristics of the analytical model in terms of absolute CBFi recovery. As the simulated SBFi was programmed to increase linearly (from $1\times10^{-6}$ to $2\times10^{-6}$ mm$^2$/s), the DL recovered CBFi waveform also exhibits a slight upward shift with increasing sample index (time). This suggests that the recovered CBFi remains partially influenced by blood flow changes in the shallow layers (i.e., not perfectly separating scalp influence), whereas the two-layer analytical fitting tends to show more stable ability to isolate the scalp flow influence. The quantitative analysis of CBFi sensitivity to both CBFi and SBFi changes is provided in Sections 3.3 and 3.4.

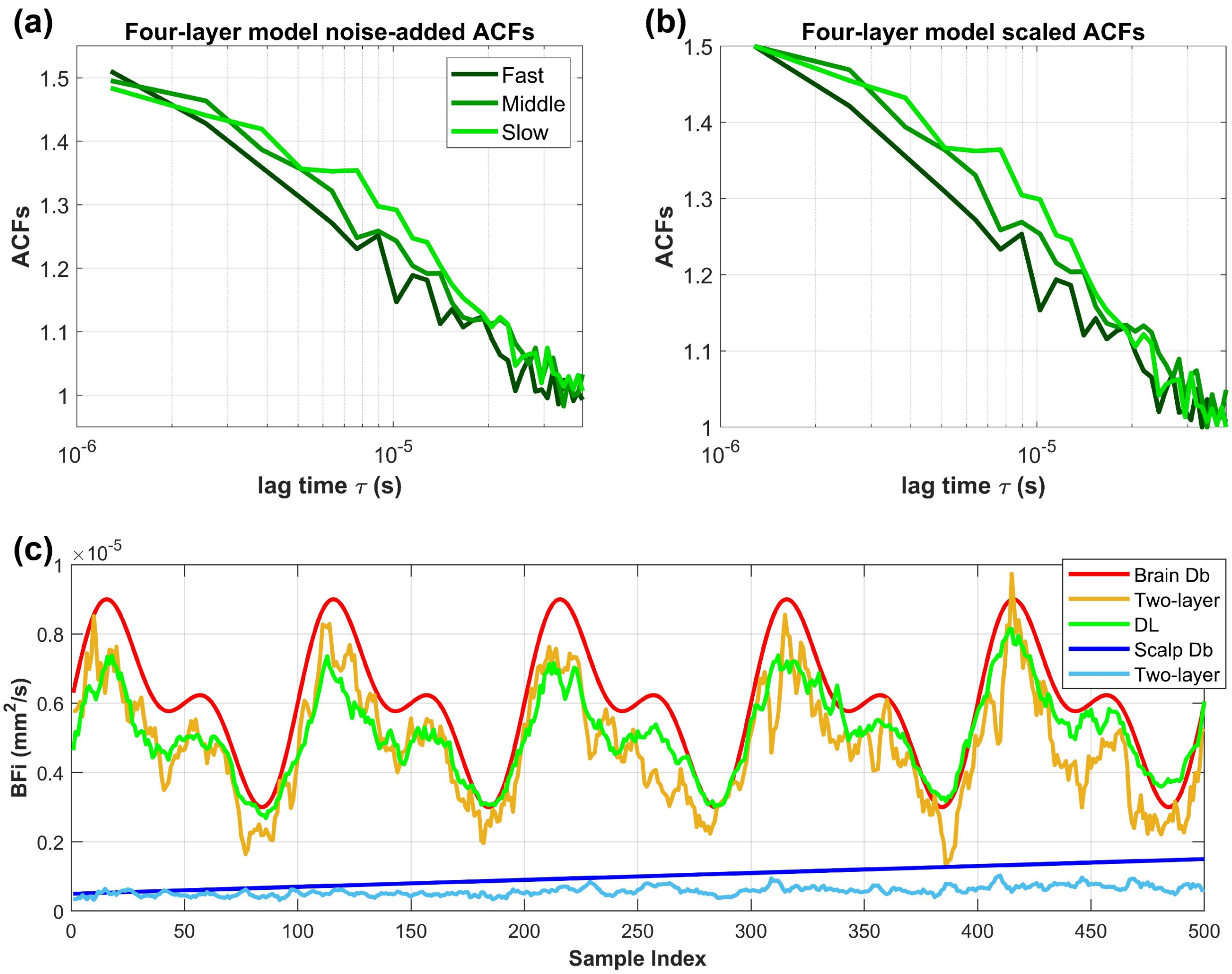

**Fig. 6** (a) Representative noise-added ACFs from the test dataset simulated with the four-layer head model. (b) Scaled DL format ACFs ready for input to the model for CBFi estimation. Note that for the two-layer analytical fitting, the used noise model has been introduced in Section 2.5. (c) Pulsatile CBFi waveform recovered from the test dataset compared with the two-layer analytical model fitting. All displayed curves were smoothed in MATLAB (function: *smooth()*, default method: moving average, span = 5). The green curve represents the CBFi waveform estimated by the DL model, whereas the yellow curve represents CBFi recovered by the two-layer analytical fitting. The red curve denotes the simulated brain $D_b$ (ground truth), the blue line corresponds to the simulated scalp $D_b$, and the light blue curve is the recovered extracerebral BFi by the two-layer analytical fitting.

Figure 7 presents the recovered rCBFi using the proposed DL model, the two-layer analytical fitting, and the relative changes in decorrelation speed obtained via single-exponential fitting. Please note that the single-exponential fitting was applied to the scaled ACFs formatted for the DL model. We have verified that this scaling has a negligible impact on the recovery of relative decorrelation speed (validation results are provided in Fig. S4, Supplementary Material).

Figure 7 shows that the two-layer analytical fitting most accurately recovers rCBFi relative to the ground truth. While both the DL model and the single-exponential fitting capture the trend of relative rCBFi changes, they tend to underestimate the magnitude of these changes. Among the two, the DL model provides a closer approximation to the ground truth than the single-exponential fitting. Additionally, compared to the two-layer analytical fitting, both the DL model and single-exponential fitting show an increasing trend in recovered relative changes as SBFi increases, suggesting that they are influenced by SBFi variations. It is worth noting that fitting the 500 data samples using the two-layer analytical model required approximately 6.7 hours (see Supplementary Material for fitting details). The quantitative analysis and comparison of these recovered relative blood flow changes is provided in Sections 3.3 and 3.4.

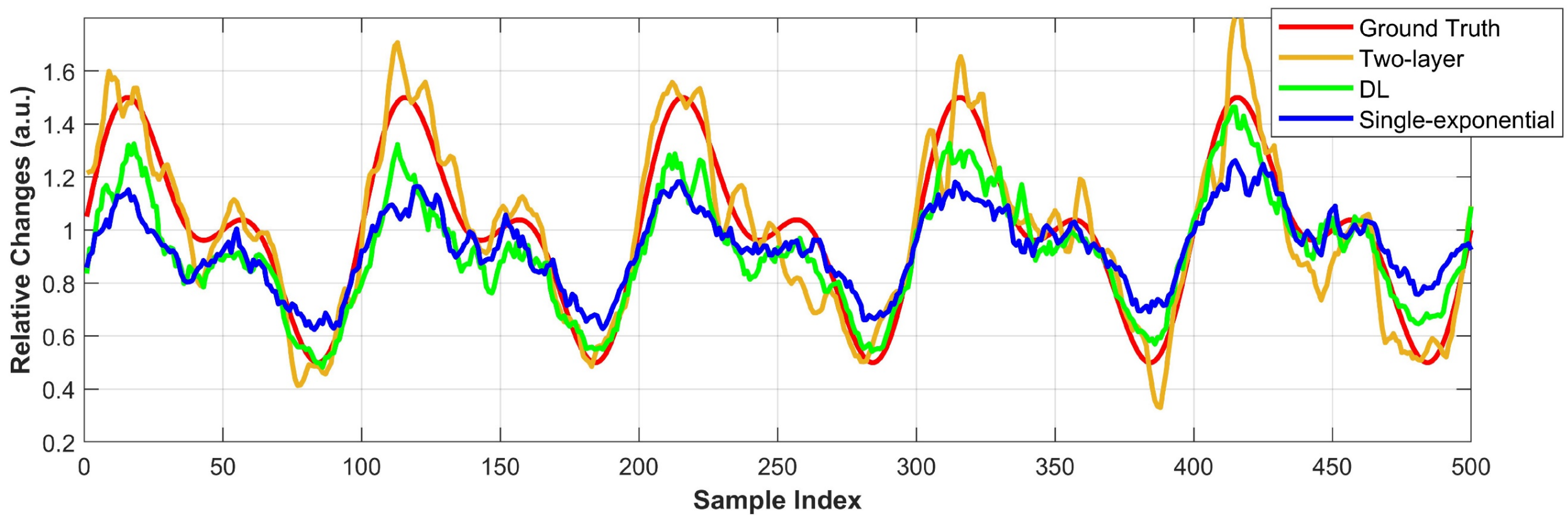

**Fig. 7** Recovered relative CBF changes using the two-layer analytical fitting (yellow line), DL model (green line) and single-exponential fitting (blue line) on the simulated dataset, with a smoothing function applied. The red line denotes the simulated ground truth.

*3.3 Model's brain sensitivity*

In this section, we quantified the model's brain sensitivity calculated using Eq. (11). Baseline CBFi was obtained using the mean value of the recovered CBFi from 1000 noise-added data. As described in Sec. 2.4, CBFi was varied by ±25% and ±50% relative to the baseline. The results were visualized using bar graphs, as shown in Fig. 8. The bar heights represent the mean of the recovered data, and the error bars indicate the standard error of the mean, estimated using nonparametric bootstrapping (1000 resamples). This approach facilitates a more statistically robust comparison of central trends across conditions by quantifying uncertainty in the mean. Fig. 8(a) shows that the proposed DL model demonstrates higher sensitivity to CBFi variations across all levels compared to single-exponential fitting, although neither method reaches full sensitivity (100%). On average, the DL model achieves a CBFi sensitivity of 86.5%, while single-exponential fitting reaches only 57.7%. In contrast, the two-layer analytical fitting exhibits superior performance, achieving an average sensitivity of 103.6%. Overall, the DL model improves CBFi sensitivity by approximately 50% compared to single-exponential fitting, although it still falls short of the accuracy provided by the two-layer analytical approach.

Fig. 8(b) presents the recovered CBF changes across different variation levels. The DL model demonstrates more accurate recovery of relative changes compared to single-exponential fitting, with an average relative error of 4.1%, versus 12.7% for the latter, as calculated using Eq. (12). Both methods tend to underestimate the true CBF changes. In contrast, the two-layer analytical fitting yields the most accurate results, with the smallest average error of 1.2%, consistent with the findings shown in Fig. 7. Additionally, in Fig. 8(b), the error bars at different $D_{b_brain}$ variation levels indicate that the DL model has a lower standard deviation than fitting when $D_{b_brain}$ is smaller than the baseline. However, the opposite trend is observed when $D_{b_brain}$ is larger than the baseline. This suggests that the proposed DL model produces more consistent estimates (i.e., lower variance) at low flow rates, but exhibits greater variability at higher flow rates compared to single-exponential fitting. This is likely due to the rapid decay of the autocorrelation function at high flow rates, which causes a loss of informative features in the measured ACF due to the limited time lag range. As a result, the model tends to underestimate flow under these conditions.

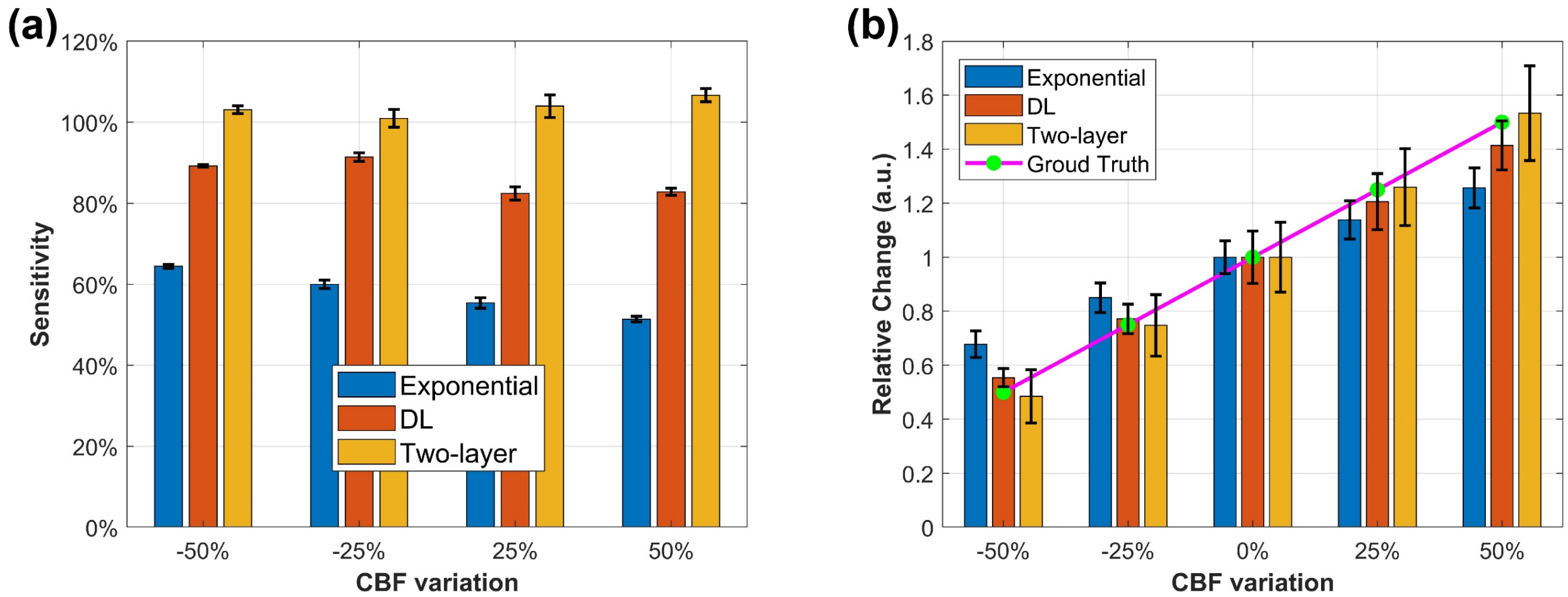

**Fig. 8** (a) and (b) Bar charts showing the recovered brain sensitivity and relative changes in response to CBFi variations are presented for single-exponential fitting, the DL model, and the two-layer analytical fitting. Note: We applied nonparametric bootstrapping to quantify the uncertainty in the estimated mean values of the sensitivity data, as the original standard deviations were relatively large, making it difficult to compare results across different methods. Specifically, we randomly resampled 1,000 data points from the original sensitivity data and calculated their mean. This process was repeated 1,000 times, resulting in a distribution of bootstrap means. The error bars in (b) represent the 25th and 75th percentiles of the recovered data, providing an interquartile range-based measure of variability.

### *3.4 Model's ability to separate extracerebral blood flow confounder*

Fig. 9(a) shows the recovered CBFi sensitivity to SBFi. The proposed DL model exhibits scalp sensitivity comparable to that of single-exponential fitting, with average sensitivities of 11.4% and 10.2%, respectively. In contrast, the two-layer analytical fitting shows an average sensitivity of -1.3%, where the negative value (as defined by Eq. 11) indicates an inverse relationship between CBFi and changes in SBFi.

For rCBFi recovery, the results are presented in Fig. 9(b). All three methods demonstrate relatively good accuracy, with average errors of 3.8% for the DL model, 3.1% for single-exponential fitting, and 0.7% for the two-layer analytical fitting, based on Eq. (12). These results suggest that both the DL model and single-exponential fitting are similarly effective at minimizing the influence of extracerebral layer changes when estimating CBF, whereas the two-layer analytical model achieves the highest accuracy but may produce a slight inverse response.

It is also worth noting that the error bars in both figures indicate the DL model exhibits a slightly larger standard deviation compared to single-exponential fitting, implying that the DL estimates are somewhat more variable in response to changes in scalp blood flow.

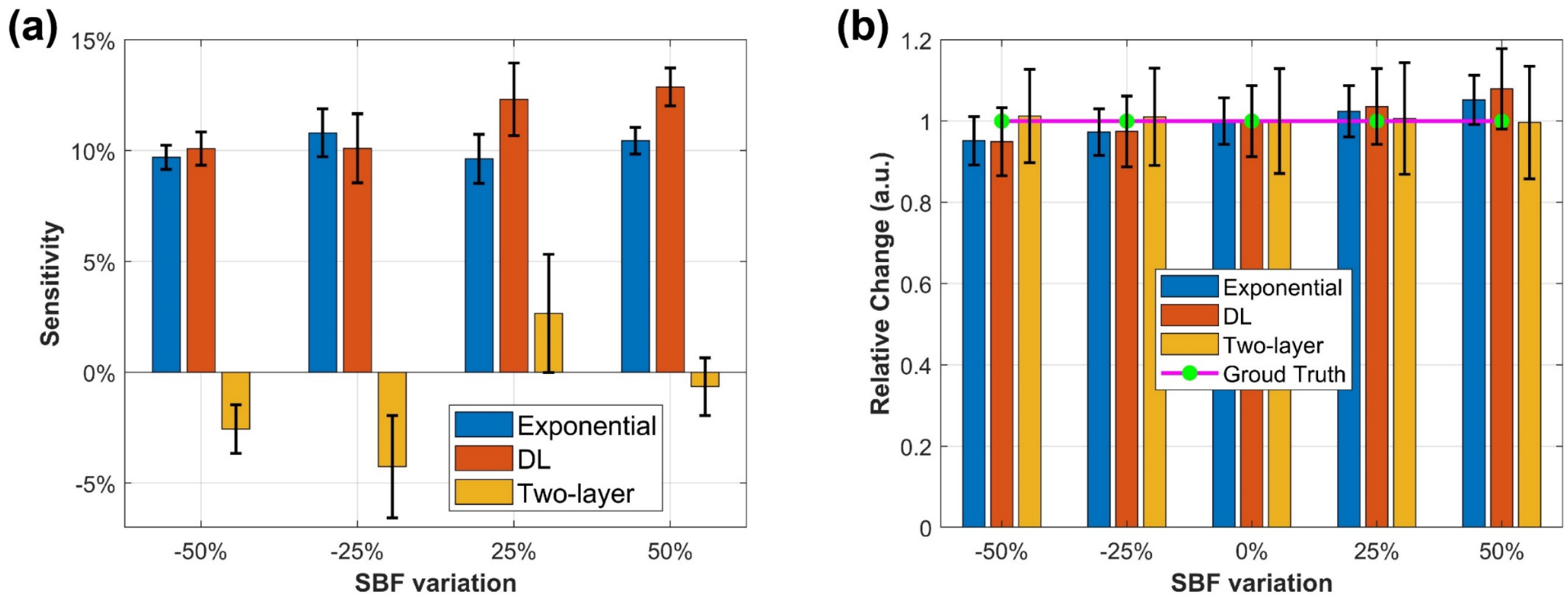

**Fig. 9** (a) and (b) Bar charts showing the recovered brain sensitivity and relative change in response to SBFi variation for single-exponential fitting, DL model, and the two-layer fitting. The error bars in (a) were calculated using the bootstrap distribution of mean values, following the same procedure used for the CBF sensitivity analysis. The error bars represent the 25th and 75th percentiles of the recovered data.

### *3.5 Model's stability test*

In this section, we compared the stability of the proposed DL model with that of the two-layer analytical model. The dataset used for the two-layer analytical fitting was the same as described in Section 2.6, and noise was added as outlined in Supplementary Material. Absolute CBFi and rCBFi errors were calculated using Eq. 13. Fig. 10 presents the absolute CBFi recovery errors for the DL method and the two-layer analytical fitting. The DL model consistently demonstrates lower CBFi errors across variations in brain layer $\mu_a$, $\mu_s'$, $L_{scalp}$, and $L_{skull}$, suggesting improved robustness. These results demonstrate that the DL model maintains good stability across different parameter settings.

In Fig. 11, the DL model shows comparable stability in rCBFi recovery relative to the two-layer analytical model, although the latter achieves slightly lower rCBFi errors. These results indicate that both methods are capable of accurately recovering rCBFi across varying tissue parameters.

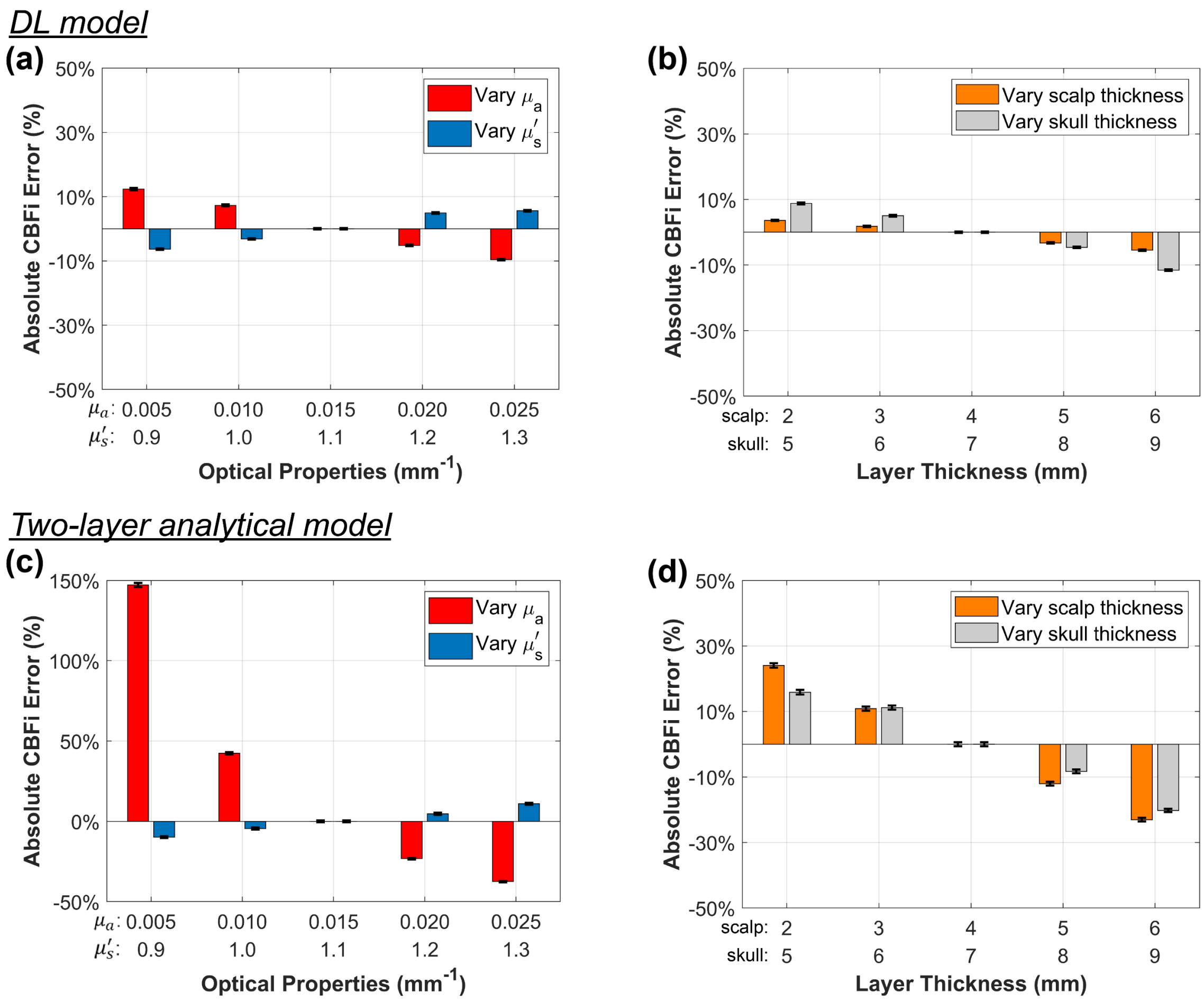

**Fig. 10** Comparison of absolute CBFi recovery between the DL method and the two-layer analytical model fitting. Bootstrapping was applied to estimate the mean CBFi errors from 1000 data points, following the same procedure described in the previous section. CBFi errors were calculated using Eq. (13). When varying one parameter, all other parameters were fixed at their baseline values ($\mu_a$= 0.015, $\mu_s'$= 1.1, $L_{scalp}$= 4 mm, and $L_{skull}$= 7 mm). These baseline values were also assumed in the two-layer analytical model fitting for consistency.

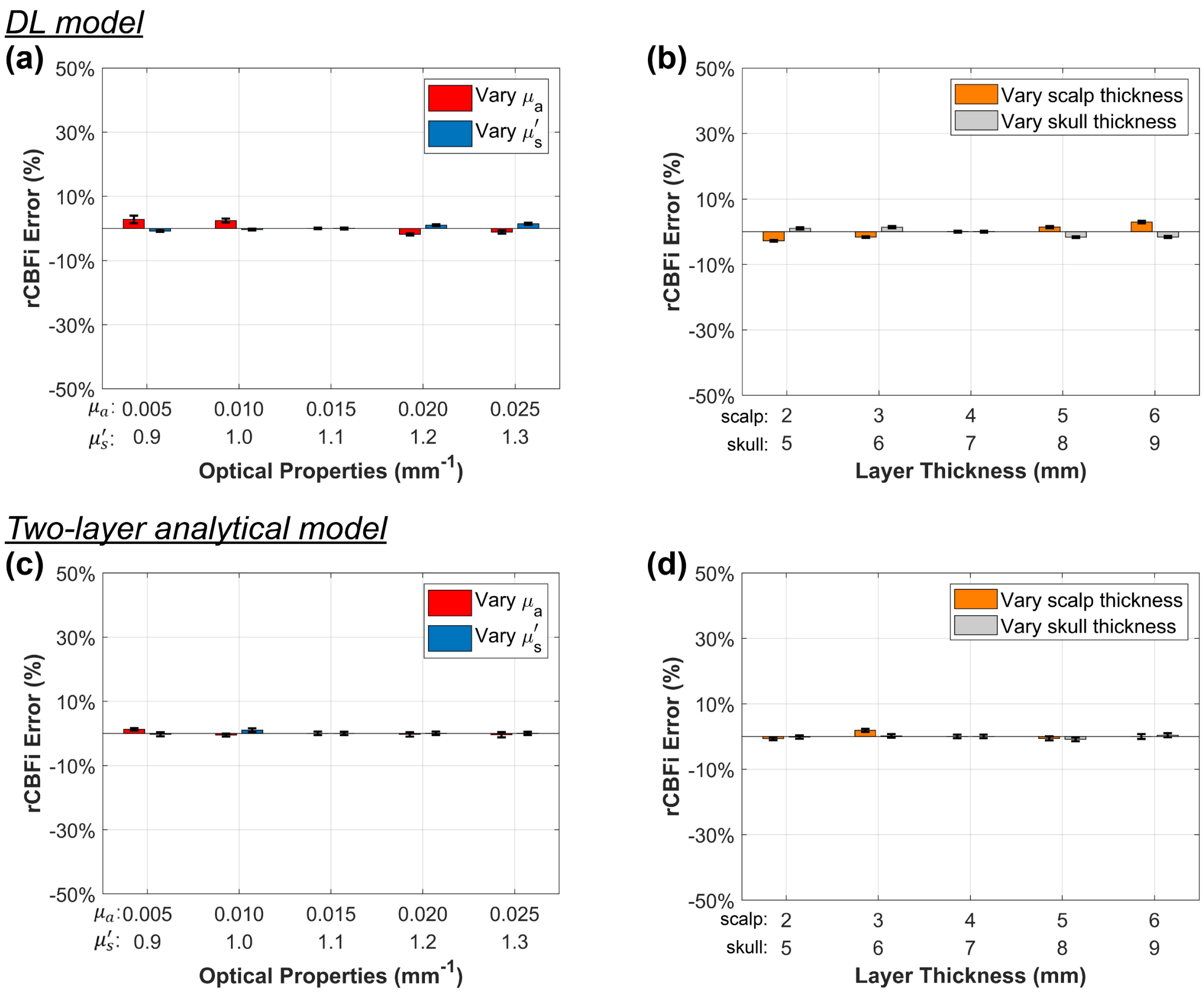

**Fig. 11** Comparison of rCBFi recovery between the DL method and the two-layer analytical model fitting. The same strategy used in Fig. 10 for absolute CBFi comparison was applied here. rCBFi was calculated based on ±50% perturbations relative to the baseline condition. The displayed rCBFi errors were computed using Eq. (13), with the errors from the two perturbation levels averaged to remove bias related to the direction of variation.

### *3.6 Baseline blood flow waveform recovery*

In this section, we evaluated the recovered cerebral perfusion waveform using the proposed DL model under the resting-state (5000 frames) in the subject. Fig. 12 presents the DL-recovered and single-exponential fitting-recovered cerebral perfusion waveforms at the baseline, we applied a third-order Butterworth low-pass filter (cutoff frequency = 0.15, normalized) with zero-phase filtering for the waveform recovery. Pearson's correlation analysis showed a strong linear relationship between the two methods ($r = 0.96$, $R^2 = 0.92$, $n = 5000$), indicating that our model closely matches the waveform recovered by the traditional curve-fitting method. The Bland-Altman plot in Fig. 12(d) further illustrates the consistency between the two methods. The mean difference is close to zero ($5.08\times10^{-5}$), and 95% of the differences lie within the limits of agreement (±1.96 SD), ranging from -0.13492 to 0.13502. These results confirm a high level of agreement and minimal systematic bias between the DL and single-exponential methods for rCBFi estimation.

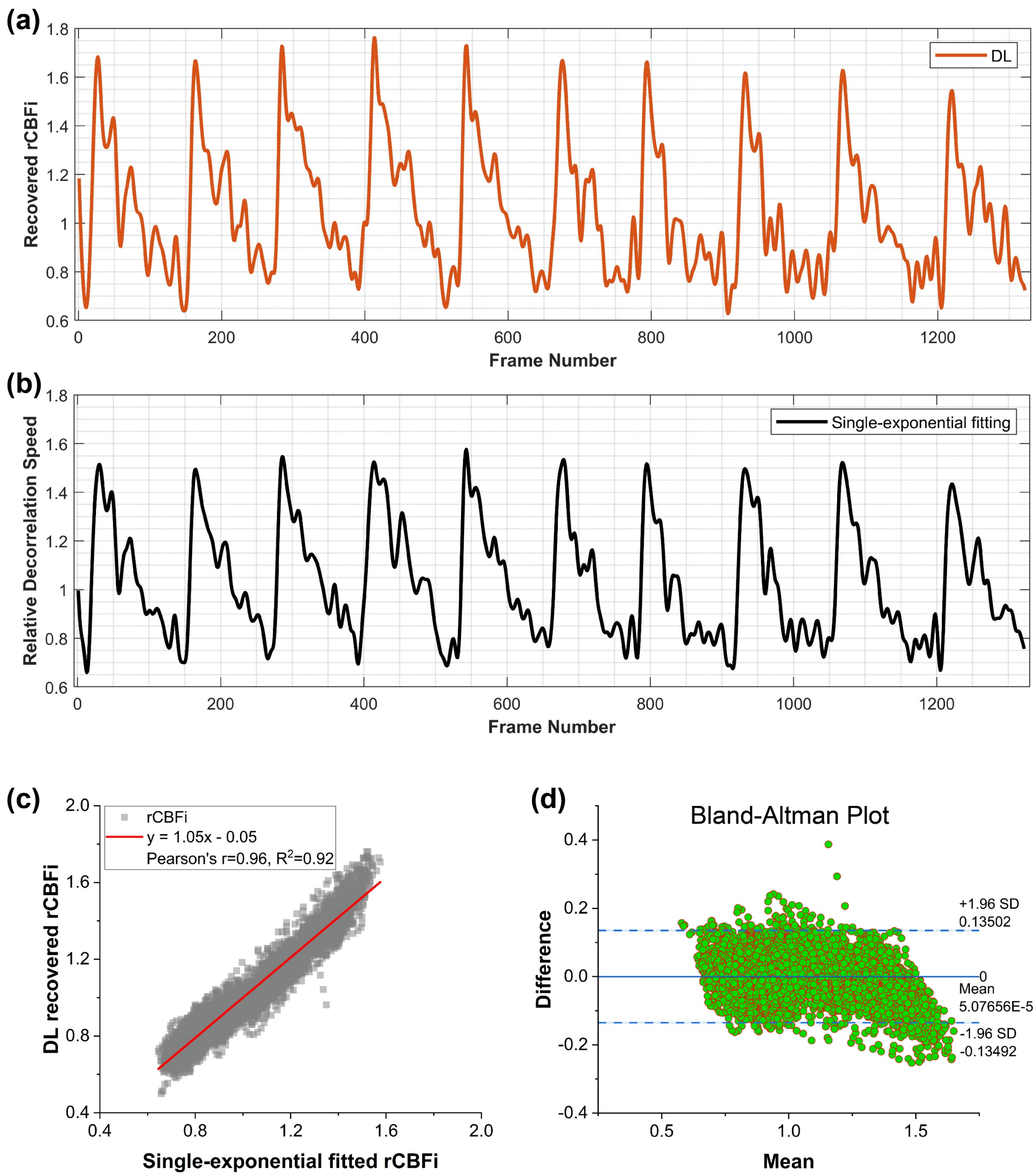

**Fig. 12** Recovered baseline cerebral perfusion waveform (30 mins before lunch). (a) and (b) The rCBFi waveform was filtered using a third-order Butterworth low-pass filter (cutoff frequency = 0.15, normalized) with zero-phase filtering (filtfilt) to reduce high-frequency noise. The mean value of recovered CBFi by DL and the mean value of recovered decorrelation speed by single-exponential fitting were taking as the baselines to calculated the relative cerebral perfusion changes, respectively. (c) Pearson's correlation analysis and (d) Bland-Altman analysis between DL model and single-exponential fitting recovered CBF waveform.

### *3.7 Repeated breath-holding tests and a lunch test on a healthy male adult*

Fig. 13 presents the results of the subject's response to breath-holding, recovered using the proposed DL model. The test was repeated three times on different days without prior practice. In Fig. 13(a), a representative example from Test 1 is shown, where the semi-transparent gray curve represents the filtered rCBFi signal (same as used in Fig.12), and the solid black line indicates the smoothed waveform obtained using a Savitzky–Golay filter (window = 1001, order = 3). The smoothed curve clearly captures a rise in rCBFi during the breath-holding phase (30-60 s), followed by a return to baseline during the recovery phase (60-90 s), consistent with the expected physiological response. Fig. 13(b) displays the rCBFi waveforms from all three repeated trials, showing cerebral blood flow increases of approximately 21.2%, 20.3%, and 17.8% at the end of the breath-holding phase for Test 1, Test 2, and Test 3, respectively. Despite day-to-day variability, all three tests consistently show an increase in rCBFi during the breath-holding period, reflecting elevated cerebral perfusion in response to hypercapnia. These results demonstrate the intra-subject repeatability of the DL model in capturing dynamic cerebral blood flow changes induced by breath-holding.

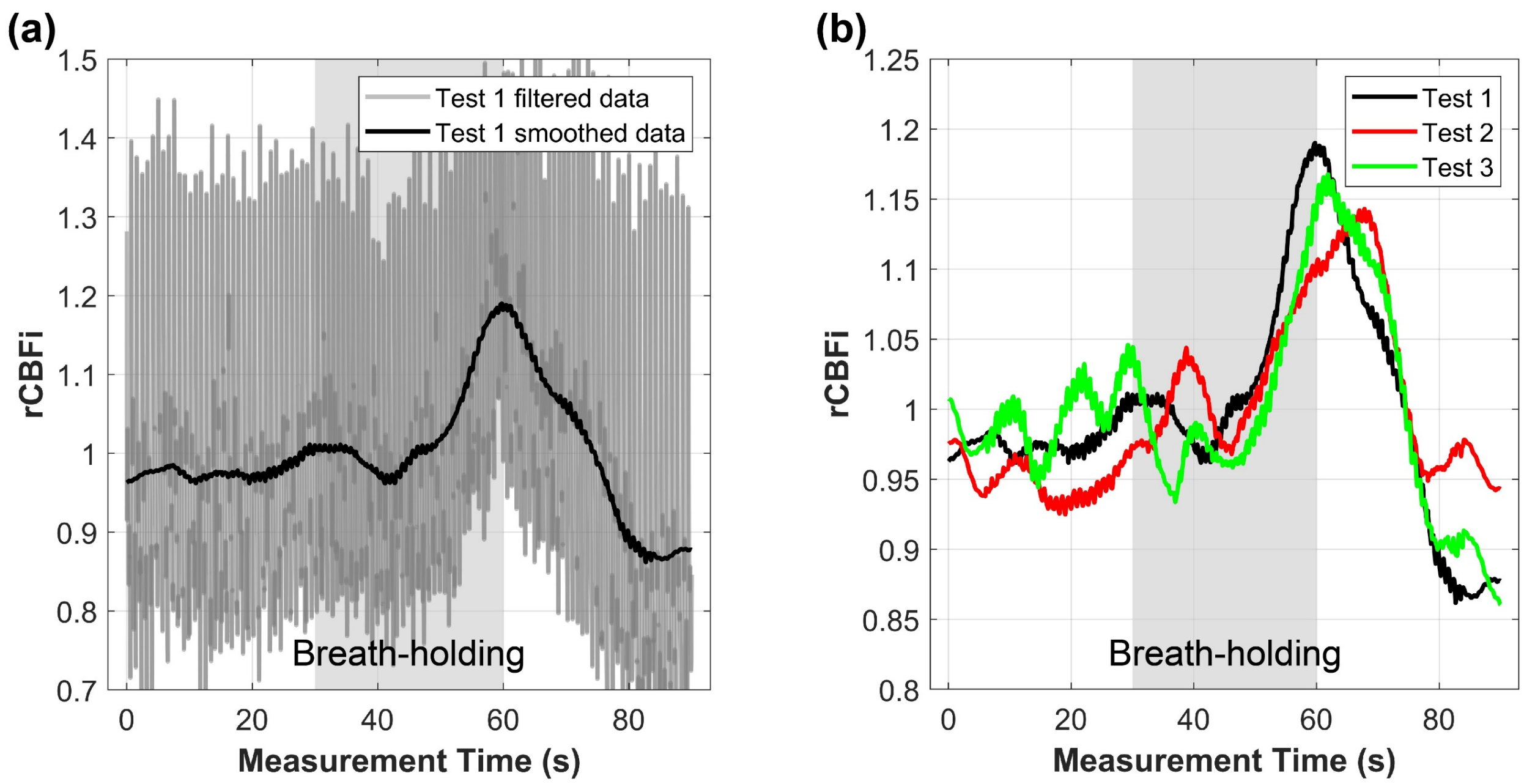

**Fig. 13** Breath-holding test results on a healthy male adult, repeated three times. (a) A representative recovered rCBFi (test 1) using our DL model, the semi-transparent gray curve is the filtered (see Fig. 12) blood flow waveform during the whole test session, the solid black curve is the smoothed rCBFi using a Savitzky–Golay filter (window = 1001, order = 3). (b) The subject's rCBFi response to breath-holding test, repeated three times (tested on different days, without practice before testing). The first 30 s is normal breath, the second 30 s is an expiratory breath hold, and the last 30 s is recovering back to normal breath.

Fig. 14 presents cerebral perfusion changes recovered by our model during the lunch test. The test conducted 30 minutes before lunch was used as the global baseline for calculating rCBFi at subsequent test phases. The results indicate a slight increase in cerebral perfusion immediately after lunch (5 minutes post-meal, light blue curves in Fig. 14), likely due to cortical stimulation from gustatory stimulation[51,52]. At 30 min, as blood flow is redirected to digestion, a short-term cerebral perfusion decrease occurred, followed by a return to baseline at 75 min (end of postprandial hyperemia[53]). Interestingly, at 120 min after lunch, a notable decrease was observed, coinciding with the subject's reported fatigue and drowsiness[54].

Additionally, we evaluated the computational efficiency of the DL model. The average processing time per 5,000 frames was 0.06 seconds, using our workstation GPU compared to 44.98 seconds for single-exponential fitting on our workstation CPU (CPU: Intel(R) Core™ i9-10900X @ 3.70

GHz; Memory: 128 GB; GPU: NVIDIA Quadro RTX 5000). This represents a 750-fold speed improvement, making our approach more suitable for neurophotonics applications where continuous monitoring and fast feedback are needed (e.g. bedside CBF tracking or neurofeedback).

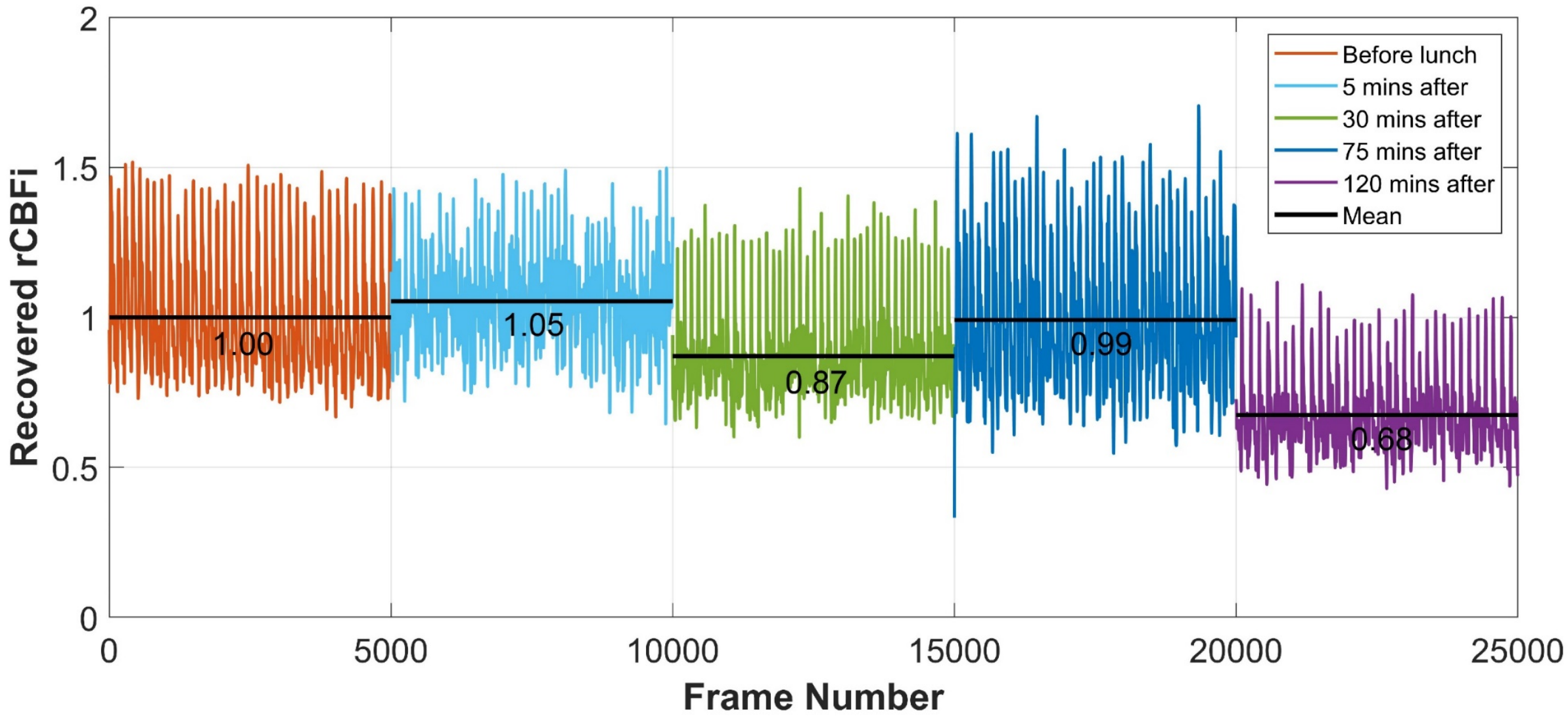

**Fig. 14** Cerebral perfusion monitoring using the proposed DL model during the lunch test. The black lines represent the mean values of the recovered rCBFi at each test phase. The mean value from the first test phase (30 minutes before lunch) was used as the global baseline for rCBFi calculation.

## 4 Discussion

We successfully implemented a DL model trained on the two-layer DCS analytical model-generated dataset with noise calculated from subject-specific baseline measurements. Our model demonstrated improved CBFi sensitivity 86.5% vs 57.7%, rCBFi error 4.1% vs 12.7% on simulated dataset, and 750× faster than single-exponential fitting method during the physiological response test. Compared with the two-layer analytical fitting, our model exhibited greater stability in absolute CBFi recovery. Notably, the two-layer analytical model showed a CBFi sensitivity exceeding 100%, indicating an overestimation of CBF changes, i.e., the recovered rCBFi fluctuated more than the ground truth (Fig. 7 and 8). In our analysis, the scalp and skull were grouped as the extracerebral layer, while the CSF and brain were grouped as the cerebral layer. This anatomical simplification may have introduced model mismatch when applying the two-layer analytical model to more complex head geometries. The impact of this grouping strategy on sensitivity and recovery accuracy will be further examined in future work.

Although our model showed excellent performance, it is worth noting that single-exponential fitting also performed well in our study, significantly enhancing CBF sensitivity and effectively minimizing the influence of the scalp layer relative to older conventional DCS setups (Fig. 9). This improvement can be attributed to advances in SPAD sensor technology, which enables early time lag detection while substantially improving the measurement SNR[18–20,55]. Single-exponential fitting, a simplified implementation of the semi-infinite analytical solution, is commonly used to characterize relative CBF changes[13]. Since the early part of the $g_2$ curve is primarily influenced by brain blood flow, while the later part is dominated by scalp blood flow, fitting only the early portion of the $g_2$ curve can enhance brain blood flow sensitivity while reducing scalp interference[6,56].

Indeed, a primary limitation of this work is also the restricted time lag range. The two-layer analytical model has demonstrated its ability to separate CBF and extracerebral blood flow (Fig. 8&9). However, during training dataset preparation, the time lag range was tailored to 1.28 μs – 39.68 $\mu s$ to match the SPAD settings, which resulted in the loss of some dynamic information from shallow layers. By training only on the early portion of the $g_2$ curve, we provided the network with less information about the slow decays from shallow flow, making it harder for the model to learn to distinguish extracerebral contributions. This limitation also explains why the DL model accurately recovers CBFi at lower flow rates but underestimates CBFi (Fig. 6) and rCBFi (Fig. 7) at higher flow rates. As BFi increases, the $g_2$ curve shifts leftward and decays faster within the limited lag window, the restricted time lag range may not capture sufficient deep tissue information. In short, there is a trade-off between brain sensitivity and the model's ability to isolate extracerebral confounders in our current system.

Moreover, as BFi decreases or $\rho$ becomes smaller, the $g_2$ curve shifts to the right, resulting in a plateau across the entire measured lag time range. Consequently, the limited lag time range constrains the applicability of our system for small $\rho$ measurements or for detecting slow blood flow (i.e., when the decorrelation time exceeds approximately 40 $\mu s$). This constraint renders the current DL model suitable only for the specific lag time range used in the system.

Regarding noise characterization in SPAD experiments, our approach provides an approximate representation of real noise. Noise levels can vary based on the detection region, source power, and inter-subjects, which means that including a broader range of noise levels would improve robustness to experimental variations. However, due to computational constraints, only three noise levels were incorporated into our dataset. Additionally, $\rho$ was fixed at 35 mm for subject-specific calibration, and the optical properties ($\mu_a$ and $\mu_s'$) in the extracerebral layer were fixed. Expanding the range of $\rho$ and sampling more extensively across other model parameters would enable broader coverage of application scenarios using the proposed method, supporting both longitudinal studies and inter-subject comparisons. We also applied a min-max scaling to both experimental and simulated data. This ensured matching ranges but could alter the shape of the ACF curve slightly (caused by noise), effectively introducing a small systematic difference between how labels were generated and how experimental data appear. In future work, more sophisticated normalization or data augmentation strategies could be explored to bridge this gap. In the breath-holding test, our method effectively captured cerebral perfusion changes induced by expiratory breath-holding. The test was repeated three times on the same subject, demonstrating strong intra-subject repeatability and highlighting the method's reliability for longitudinal monitoring. To further evaluate the system's capability for inter-subject CBF monitoring, additional breath-holding tests were conducted on two other healthy male adults. The results are presented in Figure S5 (Supplementary Material).

While the system successfully detected relative changes in cerebral perfusion during breath-holding in both subjects, the magnitude and waveform patterns of the responses varied significantly. Moreover, in contrast to the subject used to derive the noise model, the system failed to reconstruct a clear pulsatile baseline CBF waveform in these two subjects. As discussed previously, this limitation is likely due to the subject-specific nature of the noise model used during training, which may be too narrow to generalize effectively across individuals. Developing a more robust and generalizable noise model is therefore a key direction for improving the adaptability and performance of the current system.

During the lunch test, we observed slight fluctuations in recovered rCBFi curves across different test phases. Since these tests were conducted separately at different time points, variations in noise due to hardware instability may have contributed to these fluctuations. Factors such as slight differences in probe positioning, applied pressure, and detected photon count across trials could affect measurements as well. Besides, the SPAD array is highly sensitive to movement, vibration, breathing, and airflow in the test environment, all of which could introduce measurement instability. Increasing the number of noise levels in future studies may help address this issue. Moving forward, we will implement additional noise levels and optimize probe design to reduce motion and pressure variations in real-world applications. This data-driven approach can be further enhanced by incorporating additional measurement modes (e.g., multi-distance or time-resolved DCS) to provide the model with richer information for separating tissue layers.

The proposed SPAD-DCS integrated with DL method holds the ability to non-invasively monitor cerebral blood flow with high temporal resolution has broad neurophotonics implications. Competing modalities (fMRI, PET) are too slow or impractical for continuous monitoring. Traditional DCS is promising but struggled with accuracy at deep layers. The contributions of this work (layer-aware DL model + high-SNR SPAD) address those limitations head-on, bringing DCS closer to a viable neuro-monitoring tool for brain health. For example, real-time bedside monitoring in neurocritical care, neurovascular coupling studies, or augmented neuroimaging combined with functional near-infrared spectroscopy (fNIRS).

## 5 Conclusion

In this work, we demonstrated the feasibility and advantages of using a DL model based on the two-layer DCS analytical model, combined with a SPAD sensor for CBFi monitoring. The proposed DL model significantly improves CBFi sensitivity and rCBFi accuracy while exhibiting a comparable ability to an early-lag single-exponential fitting in minimizing superficial layer influence. Additionally, we applied this approach to evaluate physiological responses and demonstrated its utility in monitoring CBF changes in a healthy subject. With further hardware improvements (e.g., wider lag ranges, faster readouts), more extensive noise modeling, and expanded training datasets (including more noise levels), we anticipate even more accurate and robust performance in the future.

## Funding

This work has been funded by the Engineering and Physical Science Research Council (Grant No. EP/T00097X/1; EP/T020997/1): the Quantum Technology Hub in Quantum Imaging (QuantiC) and the University of Strathclyde.

## Disclosures

The authors declare no conflicts of interest.

## Code, Data, and Materials

Data underlying the results presented in this paper are not publicly available at this time but may be obtained from the authors upon reasonable request.

## Acknowledgments

We would like to acknowledge Dr. Qianqian Fang's input on MC simulations using MCX. Mingliang Pan and Yuanzhe Zhang would also acknowledge the support from China Scholarship Council. ATLAS was designed in a project funded by Reality Labs, Meta Platforms Inc., Menlo Park, CA 94025, USA. We are grateful to STMicroelectronics for CMOS manufacturing of the device within the University of Edinburgh Collaboration Agreement. Robert Henderson is a co-founder of Singular Photonics which is commercializing the ATLAS sensor. The company has not been involved with the paper preparation or research.

# Supplementary Material:

*1. Noise model and fitting method for the two-layer analytical model*

To add noise to the four-layer model simulated test dataset for the two-layer analytical fitting, we employed the Gaussian noise model with zero mean and a standard deviation based on the noise model proposed by Zhou *et al.* to our simulations. The standard deviation is defined as:

$$\sigma(\tau) = \sqrt{\frac{T_{\text{bin}}}{T_{\text{int}}}\left[\beta^2 \frac{(1+e^{-2\Gamma T_{\text{bin}}})(1+e^{-2\Gamma\tau}) + 2m(1-e^{-2\Gamma T_{\text{bin}}})e^{-2\Gamma\tau}}{1-e^{-2\Gamma T_{\text{bin}}}} + \frac{2\beta(1+e^{-2\Gamma T_{\text{bin}}})}{\langle n\rangle} + \frac{1+\beta e^{-\Gamma\tau}}{\langle n\rangle^2}\right]^{1/2}}, \quad (1)$$

where $\Gamma$ is the decay rate, extracted by fitting the simulated $g_2(\tau)$ curves with a single exponential decay model: $g_2(\tau) = 1 + \exp(-2\Gamma\tau)$. $T_{\text{bin}}$ is the correlator bin width, for a linear-$\tau$ strategy, it's a constant. In this paper, $T_{\text{bin}}$ = 1.28 μs, the lag time resolution in the SPAD settings. $T_{\text{int}}$ is the integration time, $m$ is the bin index, $\langle n\rangle$ is the average number of photons within the bin width $T_{\text{bin}}$ ($\langle n\rangle = IT_{\text{bin}}$, where $I$ is the detected photon count rate). We assumed a photon count rate of 10 kcps at $\rho$ = 30 mm, consistent with values reported in the previous study (Ref. 39 in the main article text). During the four-layer model simulations, we observed that the unnormalized field autocorrelation function at lag time zero exhibited the relationship $G_1(\rho = 35mm, \tau = 0)/G_1(\rho = 30mm, \tau = 0) \approx 0.33$. Based on this, we estimated a photon count rate of approximately 3.3 kcps at $\rho$ = 35 mm.

To achieve a noise level comparable to that of the measured subject-derived noise for the DL model, we set the integration time $T_{\text{int}}$=180 s. Gaussian noise with zero mean and the standard deviation defined by Eq. (1) is added to the clean $g_2$ data to simulate realistic testing conditions for the two-layer analytical fitting.

The two-layer analytical fitting was used to simultaneously extract CBFi and extracerebral BFi (EBFi) by minimizing the single penalty function defined as:

$$\chi^2 = \sum_{i=1}^{N_\tau}\left[g_2^{\text{theory}}(\tau_i, \text{CBFi}, \text{EBFi}) - g_2^{\text{simulated}}(\tau_i, D_{b_\text{brain}}, D_{b_scalp})\right]^2, \quad (8)$$

where $N_\tau$ is the number of $\tau$, and $\tau_i$ is the *i*'th delay time. We use *fminsearchbnd* function in MATLAB to minimize the penalty function. CBFi and EBFi are the fitting parameter with bounds $\in$ [$10^{-9}$, $10^{-4}$] $mm^2$/s. The optical and physiological parameters were assumed to be known before fitting, $\beta$ was assumed to be known as 0.5.

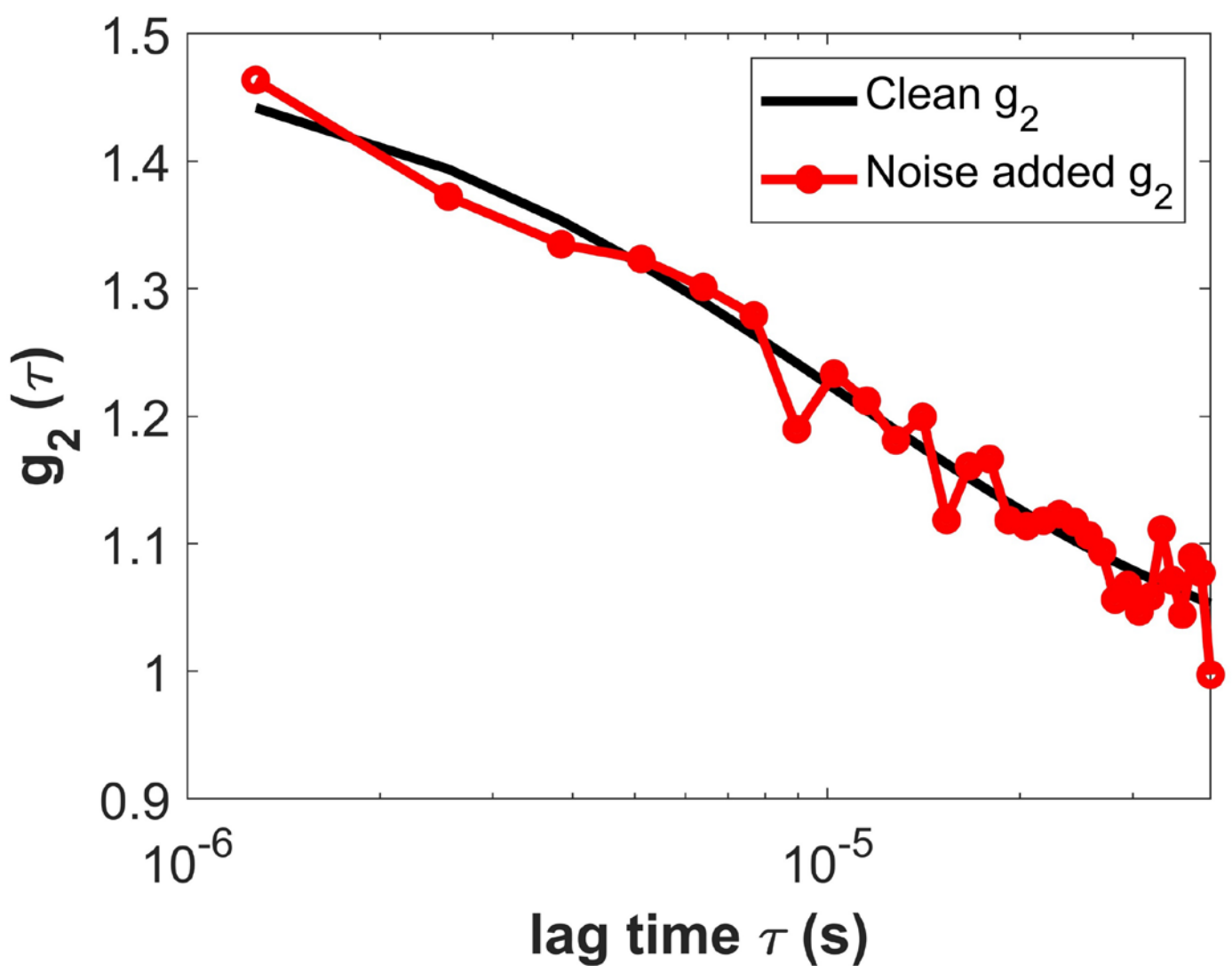

**Fig. S1** An example of clean and noise-added $g_2$ for the two-layer analytical fitting.

*2. Training loss curve*

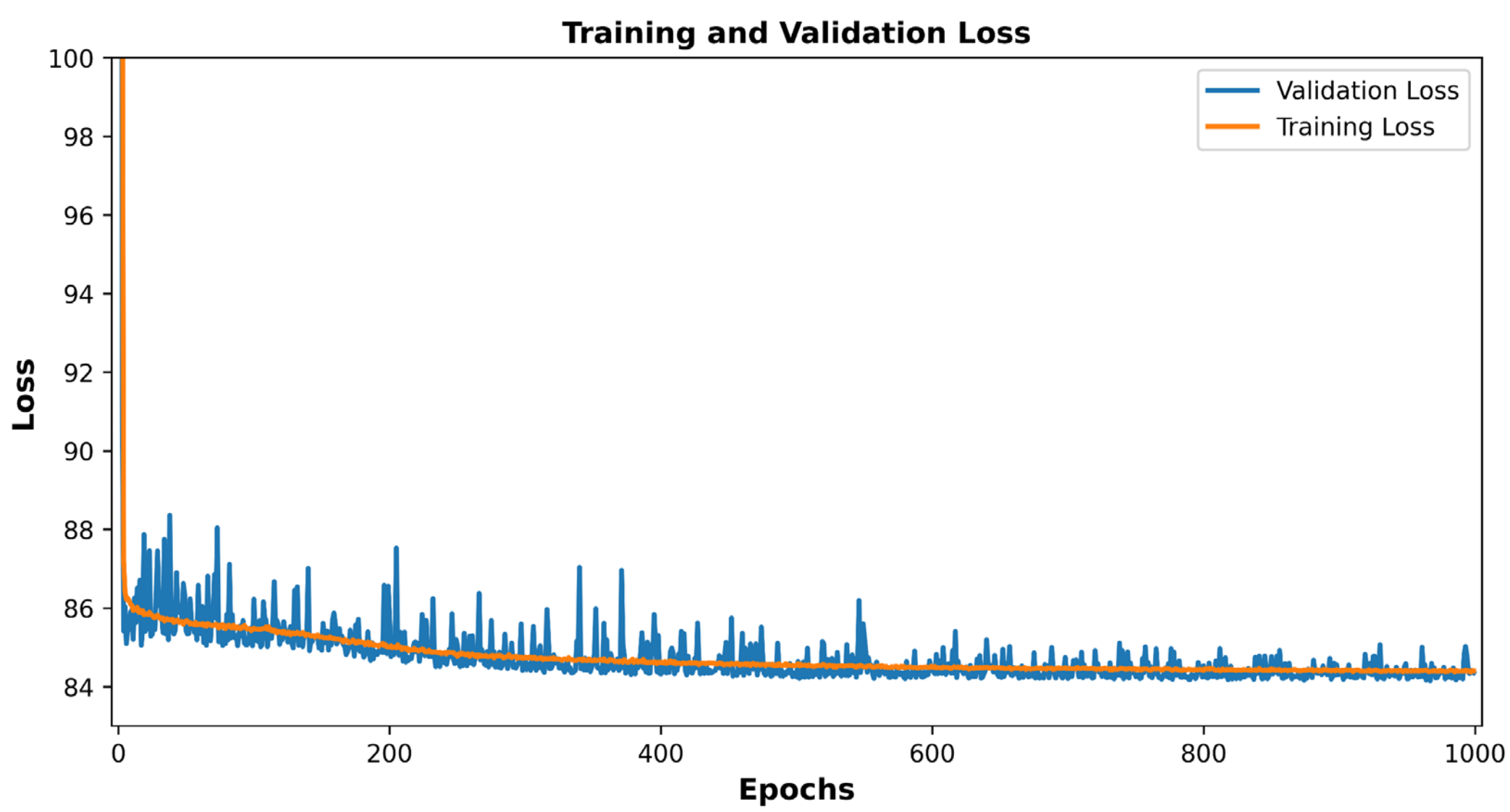

**Fig. S2** MSE training and validation loss curves over 1000 epochs. The original simulated CBFi (label) was scaled by $10^6$ to prevent slow training convergence, the training and validation losses converged to ~84.35 (in the scaled units). The total training time for 1000 epochs was 9.4 hours.

*3. Equivalence of single-exponential fitting using original and simplified decay functions*

As described in Section 2.3, we adopted a simplified function for single-exponential fitting. To validate its equivalence to the standard decay function, we compared the original form:

$$g_2(\tau) = 1 + \beta e^{-2\tau/\tau_c}, \tag{3}$$

with the simplified version:

$$g_2(\tau) = a + be^{-c\tau}. \quad (4)$$

In Eq. (3), $\tau_c$ denotes the decorrelation time. In Eq. (4), $a, b, c$ are fitting parameters, and $c$ is used to calculate the relative decorrelation speed (rCBFi). To compare the two formulations, we used the MC simulation-based test dataset described in Section 2.5, including 500 clean samples. For each sample, we generated 10 noise-added ACFs, yielding a total of 5,000 unscaled ACFs. These were fitted using both Eq. (3) and Eq. (4), and the estimated rCBFi values were compared using Pearson's correlation and Bland–Altman agreement analysis.

Figure S3(a) shows a representative recovered rCBFi waveform. The results demonstrate that both equations produce nearly identical rCBFi trends, although Eq. (4) yields slightly greater fluctuations. The linear regression between the two methods, shown in Fig. S3(b), indicates a strong correlation. Furthermore, the Bland–Altman plot in Fig. S3(c) confirms good agreement between the two approaches.

In summary, the simplified equation (Eq. 4) provides a valid alternative for single-exponential fitting. Given that the ATLAS DCS system is highly sensitive to environmental interference and probe motion, using a more flexible fitting form with fewer constraints may help reduce artifacts and more accurately reflect true physiological changes.

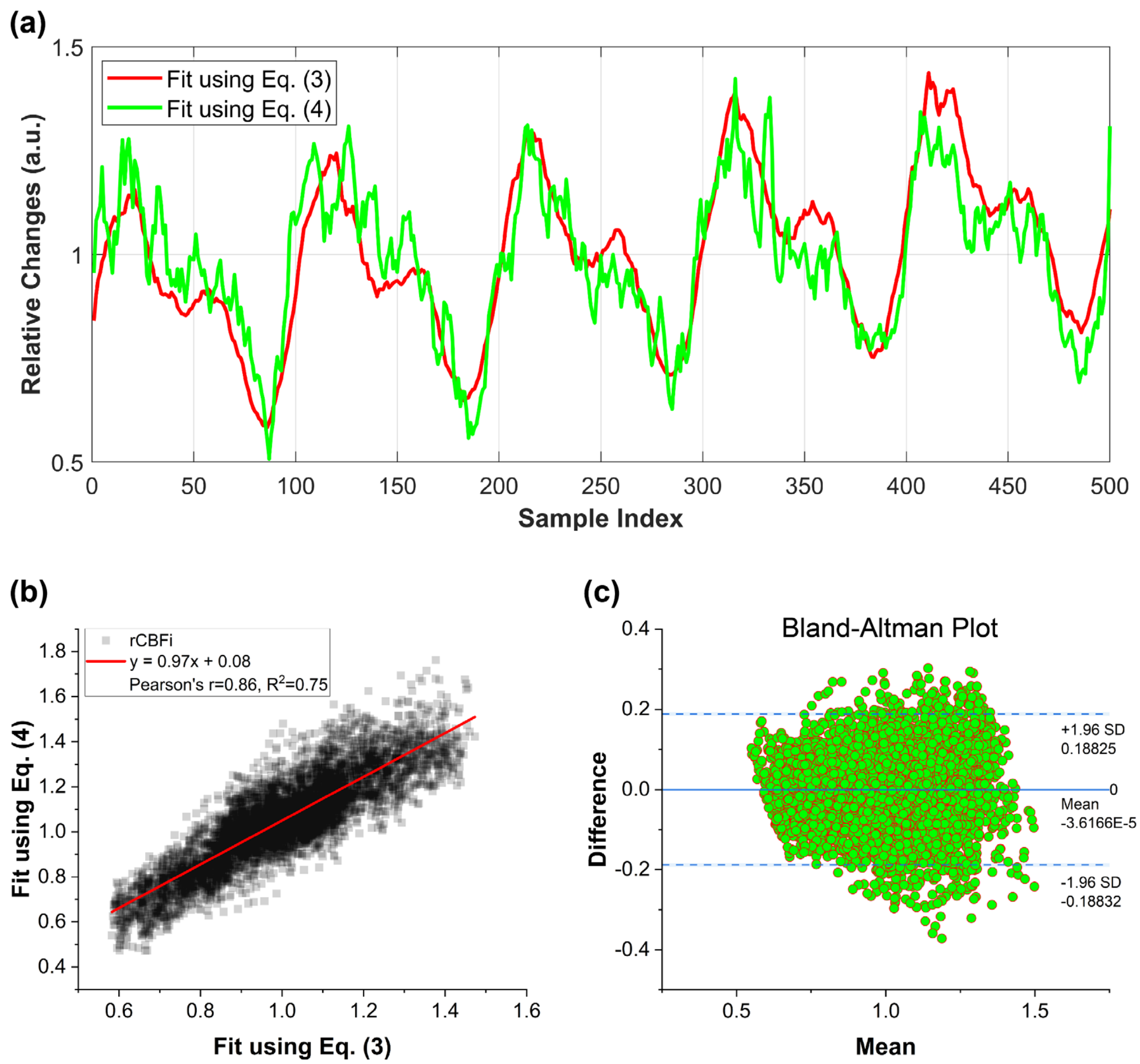

**Fig. S3** Equivalence analysis of single-exponential fitting using Eq. (3) and Eq. (4) for rCBFi recovery on unscaled ACFs. (a) Representative rCBFi waveforms recovered using Eq. (3) and Eq. (4). (b) Pearson correlation analysis of rCBFi values obtained from the two fitting methods. (c) The Bland–Altman analysis showing the distribution of differences between the two methods; among the 5,000 total samples, 258 data points lie outside the limits of agreement.

## *4. Agreement analysis of single-exponential fitting on scaled and non-scaled ACFs*

To evaluate the reliability of single-exponential fitting on the simulated test dataset, we compared its performance on both scaled and non-scaled ACFs from the baseline subject test. Fig. S4(a) shows the relative decorrelation speed estimated using both methods, which exhibit nearly identical waveforms. Fig. S4(b) presents the percentage difference between the results obtained from non-scaled and scaled ACFs, with all differences confined within ±0.5%, indicating minimal impact from the scaling process.

As shown in Fig. S4(c), the fitted values from both methods align closely along the identity line (y = x), confirming strong agreement. Additionally, the Bland–Altman plot in Fig. S4(d) further supports this finding. The differences are centered around zero with a mean bias of $-2.28\times10^{-7}$ and 95% of the data falling within the limits of agreement (±1.96 SD, or approximately ±0.00113).

These results demonstrate that the scaling of ACFs has a negligible effect on the outcome of single-exponential fitting, confirming its robustness under both input formats.

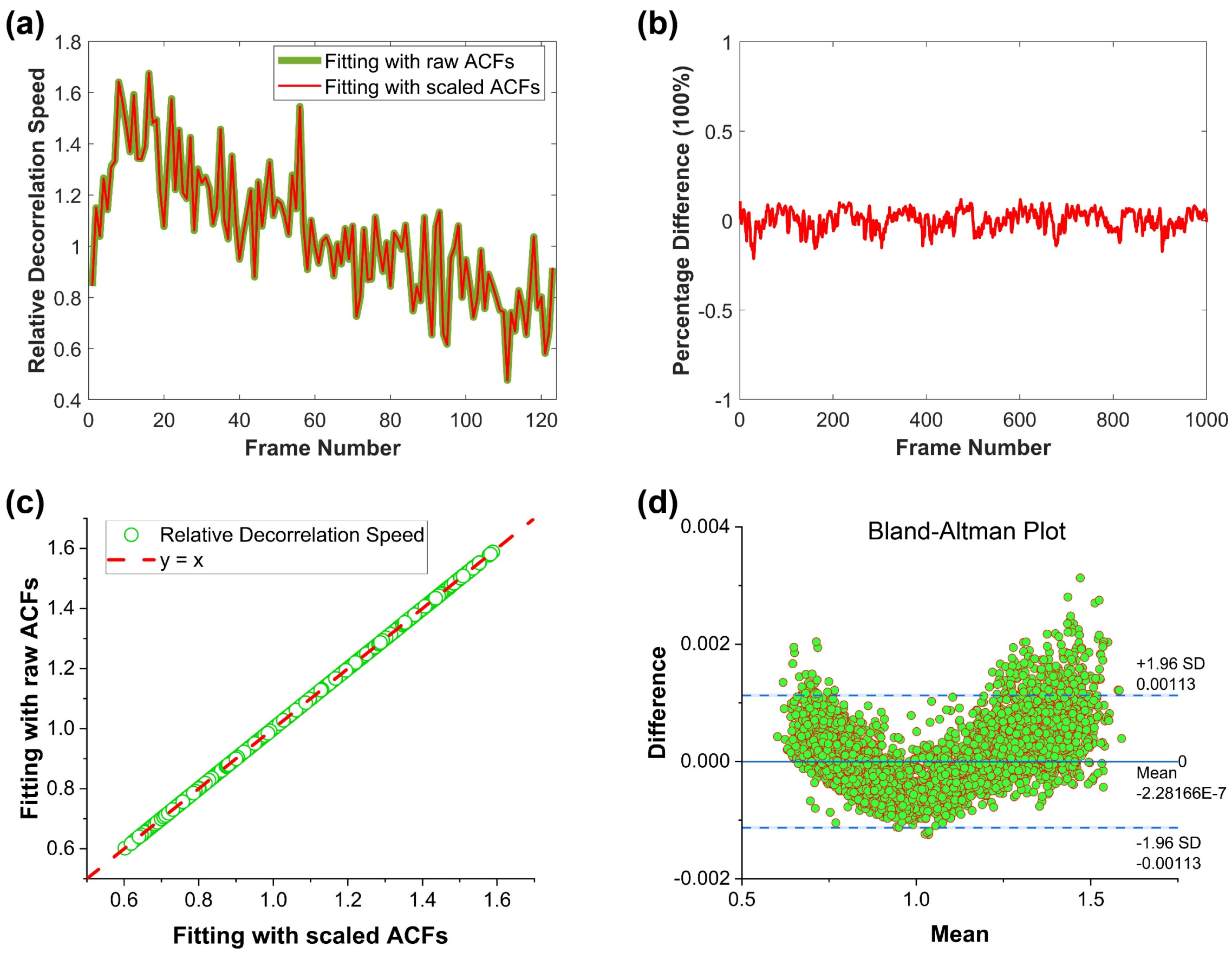

**Fig. S4** Validation of the agreement between single-exponential fitting applied to scaled and non-scaled ACFs. (a) Recovered relative decorrelation speed using non-scaled and scaled ACFs during one pulse period. (b) Percentage difference between the fitted relative decorrelation speeds from the two input formats. (c) Scatter plot comparing the results from both methods against the identity line (y = x), indicating strong agreement. (d) Bland–Altman analysis showing the distribution of differences; out of 5000 total samples, 236 data points fall outside the limits of agreement.

### *5. Inter-subject test evaluation*

To evaluate the feasibility of using our system for CBF monitoring across different subjects, we conducted additional breath-holding tests on another two healthy male participants (Subject B: 28 years old, and Subject C: 25 years old). The results, presented in Fig. S5, show that the recovered CBF waveforms for Subjects B and C differ from one another and are less distinguishable compared to the responses observed in the primary subject. Although the system successfully captured breath-hold-induced increases in CBF for both subjects, the magnitude and pattern of the responses varied considerably (27.4% rCBFi increase for Subject B, and 5.4% increase for Subject C at the end of breath-holding).

A likely explanation for this variation is that the model was trained using a subject-specific noise profile, which may be too narrow in scope to generalize effectively to data from other individuals. We recognize this as a main limitation of the current system. In future work, we aim to improve the model's generalizability by developing more robust training dataset generation strategies and incorporating a more comprehensive, subject-independent noise model.

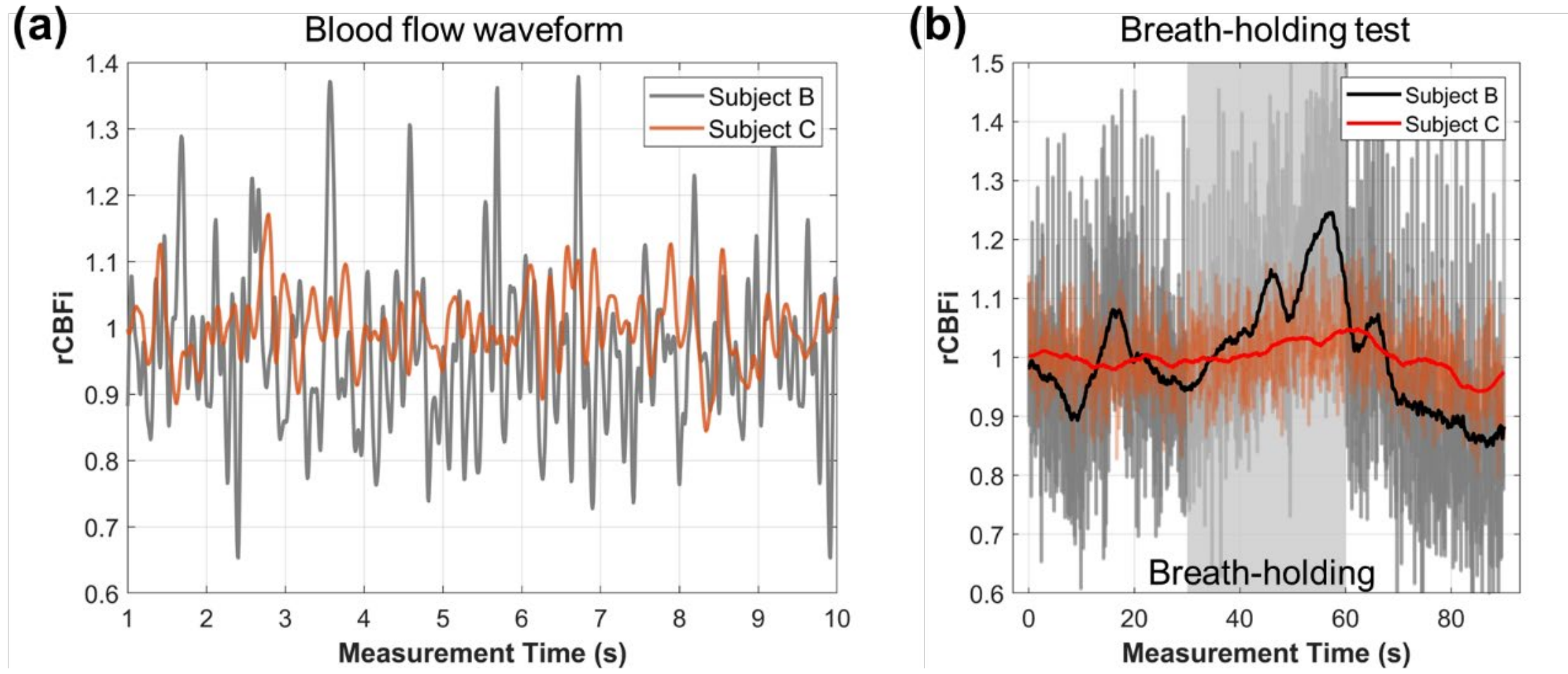

**Fig. S5** Inter-subject test results. (a) Blood waveform of the first 10 seconds tested on Subject B (28-year male adult) and C (25-year male adult). (b) Breath-holding test on Subject B and C.